\documentclass[]{aastex631}

\usepackage{color}

\shorttitle{Optical Variability of Blazars in the Tomo-e Gozen Northern Sky Transient Survey}
\shortauthors{Zhang et al.}

\graphicspath{{./}{}}

\begin{document}

\title{Optical Variability of Blazars in the Tomo-e Gozen Northern Sky Transient Survey}

\author{TianFang Zhang}
\affiliation{Department of Astronomy, Graduate School of Science, The University of Tokyo, 7-3-1, Hongō, Bunkyo City, Tokyo, 113-8654, Japan}
\affiliation{Institute of Astronomy, Graduate School of Science, The University of Tokyo, 2-21-1 Osawa, Mitaka, Tokyo, 181-0015, Japan}
\affiliation{National Astronomical Observatory of Japan, National Institutes of Natural Science, 2-21-1 Osawa, Mitaka, Tokyo, 181-8588, Japan}

\author{Mamoru Doi}
\affiliation{Institute of Astronomy, Graduate School of Science, The University of Tokyo, 2-21-1 Osawa, Mitaka, Tokyo, 181-0015, Japan}
\affiliation{National Astronomical Observatory of Japan, National Institutes of Natural Science, 2-21-1 Osawa, Mitaka, Tokyo, 181-8588, Japan}
\affiliation{
Research center for the early universe School of Science Bldg. No.4, The University of Tokyo, 7-3-1, Hongō, Bunkyo City, Tokyo, 113-0033, Japan}

\author{Mitsuru Kokubo}
\affiliation{National Astronomical Observatory of Japan, National Institutes of Natural Science, 2-21-1 Osawa, Mitaka, Tokyo, 181-8588, Japan}

\author{Shigeyuki Sako}
\affiliation{Institute of Astronomy, Graduate School of Science, The University of Tokyo, 2-21-1 Osawa, Mitaka, Tokyo, 181-0015, Japan}
\affiliation{UTokyo Organization for Planetary Space Science, The University of Tokyo, 7-3-1, Hongō, Bunkyo City, Tokyo, 113-8654, Japan}
\affiliation{Collaborative Research Organization for Space Science and Technology, The University of Tokyo, 7-3-1, Hongō, Bunkyo City, Tokyo, 113-8654, Japan}

\author{Ryou Ohsawa}
\affiliation{National Astronomical Observatory of Japan, National Institutes of Natural Science, 2-21-1 Osawa, Mitaka, Tokyo, 181-8588, Japan}

\author{Nozomu Tominaga}
\affiliation{National Astronomical Observatory of Japan, National Institutes of Natural Science, 2-21-1 Osawa, Mitaka, Tokyo, 181-8588, Japan}
\affiliation{Astronomical Science Program, Graduate Institute for Advanced Studies, SOKENDAI, 2-21-1 Osawa, Mitaka, Tokyo, 181-8588, Japan}
\affiliation{Department of Physics, Faculty of Science and Engineering, Konan University, 8-9-1 Okamoto, Kobe, Hyogo, 658-8501, Japan}

\author{Masaomi Tanaka}
\affiliation{Astronomical Institute, Tohoku University, Aoba, Sendai, 980-8578, Japan}
\affiliation{Division for the Establishment of Frontier Sciences, Organization for Advanced Studies, Tohoku University, Sendai, 980-8577, Japan}

\author{Yasushi Fukazawa}
\affiliation{Department of Physics, Hiroshima University, 1-3-1 Kagamiyama, Higashi-Hiroshima, 739-8526, Japan}

\author{Hidenori Takahashi}
\affiliation{Kiso Observatory, Institute of Astronomy, Graduate School of Science, The University of Tokyo, Mitake 10762-30, Kiso-machi, Kiso-gun, Nagano, 397-0101, Japan}
\author{Noriaki Arima}
\affiliation{Institute of Astronomy, Graduate School of Science, The University of Tokyo, 2-21-1 Osawa, Mitaka, Tokyo, 181-0015, Japan}
\author{Naoto Kobayashi}
\affiliation{Kiso Observatory, Institute of Astronomy, Graduate School of Science, The University of Tokyo, Mitake 10762-30, Kiso-machi, Kiso-gun, Nagano, 397-0101, Japan}
\author{Ko Arimatsu}
\affiliation{Hakubi Center / Astronomical Observatory, Graduate School of Science, Kyoto University, Kitashirakawa-Oiwakecho, Sakyo-ku, Kyoto, 606-8502, Japan}
\author{Shin-ichiro Okumura}
\affiliation{Japan Spaceguard Association, Bisei Spaceguard Center, 1716-3 Okura, Bisei, Ibara, Okayama, 714-1411, Japan}
\author{Sohei Kondo}
\affiliation{Kiso Observatory, Institute of Astronomy, Graduate School of Science, The University of Tokyo, Mitake 10762-30, Kiso-machi, Kiso-gun, Nagano, 397-0101, Japan}
\author{Toshihiro Kasuga}
\affiliation{National Astronomical Observatory of Japan, National Institutes of Natural Science, 2-21-1 Osawa, Mitaka, Tokyo, 181-8588, Japan}
\author{Yuki Mori}
\affiliation{Kiso Observatory, Institute of Astronomy, Graduate School of Science, The University of Tokyo, Mitake 10762-30, Kiso-machi, Kiso-gun, Nagano, 397-0101, Japan}
\author{Yuu Niino}
\affiliation{Kiso Observatory, Institute of Astronomy, Graduate School of Science, The University of Tokyo, Mitake 10762-30, Kiso-machi, Kiso-gun, Nagano, 397-0101, Japan}

\begin{abstract}
We studied the optical variability of 241 BL Lacs and 83 flat-spectrum radio
quasars (FSRQ) from the 4LAC catalog using data from the Tomo-e Gozen Northern
Sky Transient Survey, with $\sim$ 50 epochs per blazar on average. We excluded
blazars whose optical variability may be underestimated due to the influence of
their host galaxy, based on their optical luminosity ($L_O$). FSRQs with
$\gamma$-ray photon index greater than 2.6 exhibit very low optical
variability, and their distribution of standard deviation of repeated
photometry is significantly different from that of the other FSRQs (KS test P
value equal to $5 \times 10^{-6}$ ). 
Among a sample of blazars at any particular cosmological epoch,
those with lower $\gamma$-ray luminosity
($L_\gamma$) tend to have lower optical variability,
and those FSRQs with $\gamma$-ray photon index greater than 2.6 tend to have
low $L_\gamma$. 
We also measured the structure
function of optical variability 
and found that the amplitude of the structure function for FSRQs is higher
than previously measured and higher than that of BL Lacs at multiple time
lags. Additionally, the amplitude of the structure function of FSRQs with high
$\gamma$-ray photon index is significantly lower than that of FSRQs with low
$\gamma$-ray photon index. The structure function of FSRQs of high
$\gamma$-ray photon index shows a characteristic timescale of more than 10 days,
which may be the variability timescale of the accretion disk. In summary, we
infer that the optical component of FSRQs with high $\gamma$-ray photon index
may be dominated by the accretion disk.

\end{abstract}


\section{Introduction} \label{sec:intro}
Blazars are a type of active galactic nucleus (AGN) characterized by a
relativistic jet pointed directly at the
observer\citep{blandford1979relativistic}. As one of the most powerful
phenomena in the universe, blazars have garnered significant attention in the
field of extra-galactic astronomy. Jets of blazars are responsible for
producing a wide range of observable energies, the majority of which are in the
form of non-thermal radiation. The non-thermal radiation 
includes synchrotron radiation, which occurs across radio to X-ray frequencies,
as well as inverse Compton scattering, which occurs from X-ray to
$\gamma$-ray frequencies \citep{konigl1981relativistic}.

Blazars can be classified into two primary intrinsic types, namely
low-ionization (BL Lacs) and high-ionization (FSRQs, predominantly beamed
Fanaroff–Riley class II sources) (\citep{giommi2012simplified};
\citep{fanaroff1974morphology}). The distinction between BL Lacs and FSRQs is
primarily determined by the presence or absence of emission lines in their
optical-IR spectra. Specifically, BL Lacs do not display prominent emission
lines in their spectra, whereas FSRQs do display prominent emission lines in
their spectra.

In a recent study, \citep{ghisellini2017fermi} examined the dependence of the
Spectral Energy Distribution (SED) on $\gamma$-ray luminosity ($L_\gamma$)
using Fermi/LAT and other data. The study found that FSRQs exhibit a similar
SED across a wide range of $L_\gamma$, while BL Lacs display decreasing synchrotron
peak frequencies with increasing $L_\gamma$. Furthermore, the nature of blazars
at various wavelengths has been discussed by several studies
(\citep{richards2011blazars}, \citep{landi2015swift},
\citep{acciari2021multiwavelength}, \citep{kaur2021classifying},
\citep{kerby2021x}). 

Studying the optical variability of blazars is essential for understanding
their nature.
\citep{clements2003pks}
monitored the intra-night optical variability of FSRQ PKS 0736+017 and observed
a dramatic flare, with the source brightening by 1.3 magnitudes ($R$ band) in
just 2 hours. In a study comparing the intra-night optical variability rates of
blazars, \citep{bachev2012nature} found indications for the presence of
quasi-periodic micro-oscillations with periods of about an hour. Similarly,
\citep{rani2013radio} monitored the intra-night multi-wavelength luminosity of
BL Lac object S50716+714 
over a period of four years
and identified a long-term variability
trend (about 350 days) and a shorter time scales (about 60 days).
\citep{pandey2020optical} studied the optical variability of BL Lac 1ES
0806+524 in three bands ($VRI$) using high cadence observations for 153 nights
during 2011$\sim$2019, and found a small but significant variation in both $V$
band and $R$ band light curves in only one night. In a recent study,
\citep{goyal2021optical} examined the intra-night optical variability of 9 BL Lacs
and 5 FSRQs, carrying out a Fourier transform on their light curves, and
obtained their power spectral densities. They found that most of the power
spectral densities were well fit by simple power-laws with slopes ranging from
1.4 to 4.0, with no significant difference between BL Lacs and FSRQs. These
studies highlight the diversity of the optical variability of blazars which may
give insights into the physical processes driving their emission.

The optical variability of FSRQs is a topic of much debate among researchers.
One viewpoint is that FSRQs, dominated by accretion disks, should exhibit smaller
optical variability than jet-dominated BL Lacs \citep{wiita2005accretion}.
Conversely, others suggest that FSRQs are a mixture of various types, including
optically violently variable quasars, highly polarized quasars, core-dominated
quasars, and superluminal motion sources, with some objects exhibiting optical
variation of up to 50$\%$ in one day (\citep{penston1970optical};
\citep{fan2005optical}). \citep{bonning2012smarts} conducted observations on 9
FSRQs and revealed that in multiple bands ($BVRJ$), the fractional variability
amplitude of 3C 273 was one magnitude lower than that of other FSRQs, providing
support for the aforementioned bipolarity.

In a study by \citep{EmmanoulopoulosBauer_2009}, the statistical variability of
blazars, which included 276 FSRQs and 86 BL Lacs, and nearly 23000 quasars were
analyzed using data from the Palomar-QUEST Survey
\citep{djorgovski2008palomar}. The authors investigated the overall variability
of BL Lacs and FSRQs using the structure function and found that different
objects exhibit varying degrees of variability details on timescales up to a
few years.

Utilizing the Tomo-e Gozen Northern Sky Transient Survey data, we have
investigated the optical variability of blazars in this work. 
Compared to the previous work of
\citep{EmmanoulopoulosBauer_2009},
our
study has a larger ample size (with redshift) and 
a larger number of measurements per object,
with an average of $\sim$50 epochs per blazar compared to an average of 6
epochs in \citep{EmmanoulopoulosBauer_2009}. Moreover, we
have used the latest catalog, the 4LAC catalog \citep{ajello2020fourth}, to
classify blazars.

In Section \ref{section:Data}, we provide a concise description of the data
used in this study. Our investigation into the variability of blazars,
utilizing the standard deviation of repeated photometry and the structure
function, is presented in Section \ref{section:Variabilit Analysis}. In Section
\ref{Discussion}, we discuss the differences between this study and previous
studies and outline its limitations. Our conclusions are summarized in Section
\ref{Summary}. The calculations and interpretations presented in this work are
based on a cosmological model defined as a Flat Lambda Cold Dark Matter
(\(\Lambda\)CDM) cosmology. Specifically, we adopt a flat universe assumption
with Hubble constant \(H_0 = 70\, \text{km s}^{-1} \text{Mpc}^{-1}\) and a
matter density parameter \(\Omega_m = 0.3\).

\clearpage

\clearpage
\section{Method}
\label{section:Data}
\subsection{Tomo-e Gozen observation of Fermi Blazar}
\begin{figure}[ht]
\centering
\gridline{\fig{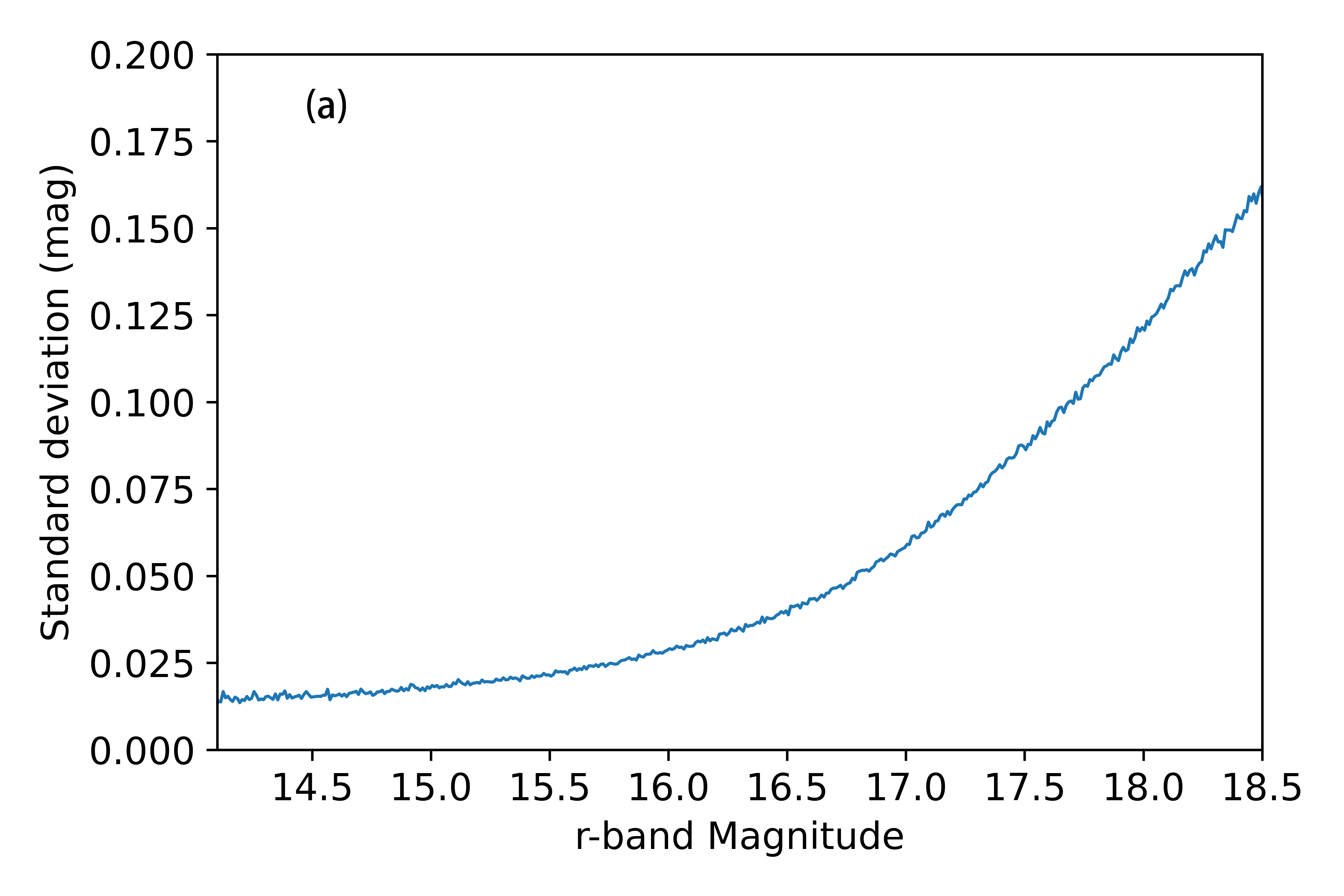}{0.48\textwidth}{(a)} 
         \fig{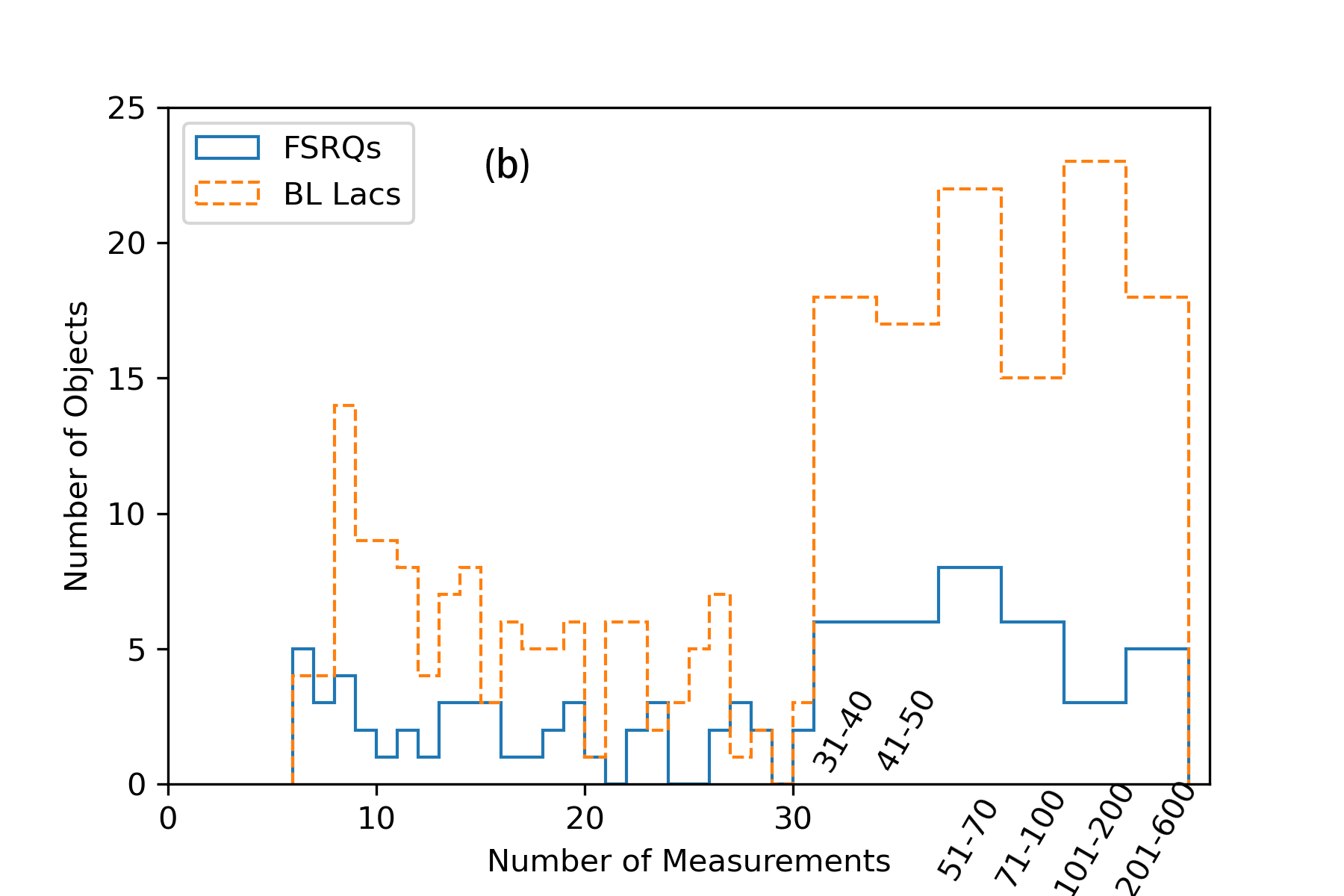}{0.5\textwidth}{(b)}
         }
\gridline{\fig{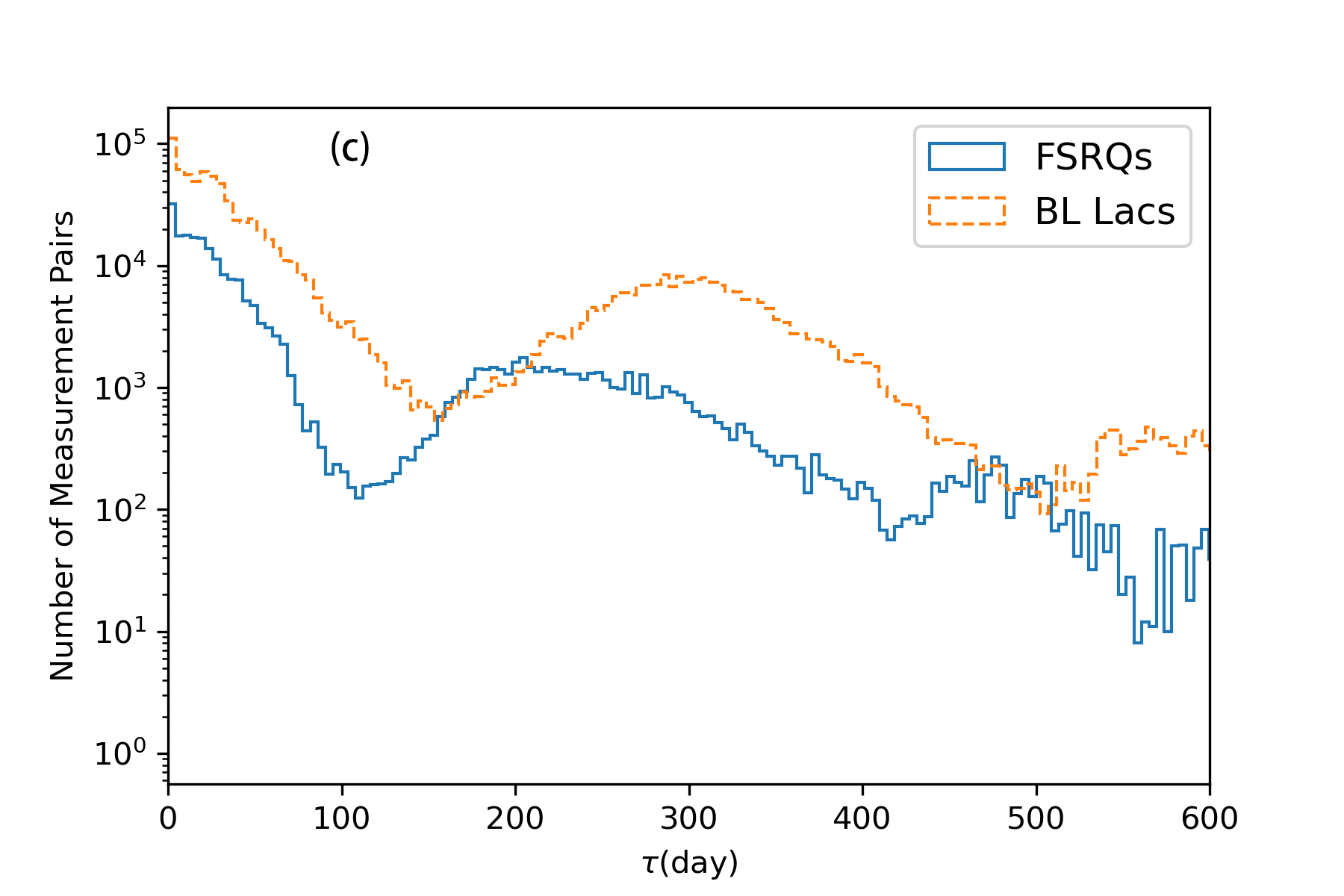}{0.5\textwidth}{(c)}
         \fig{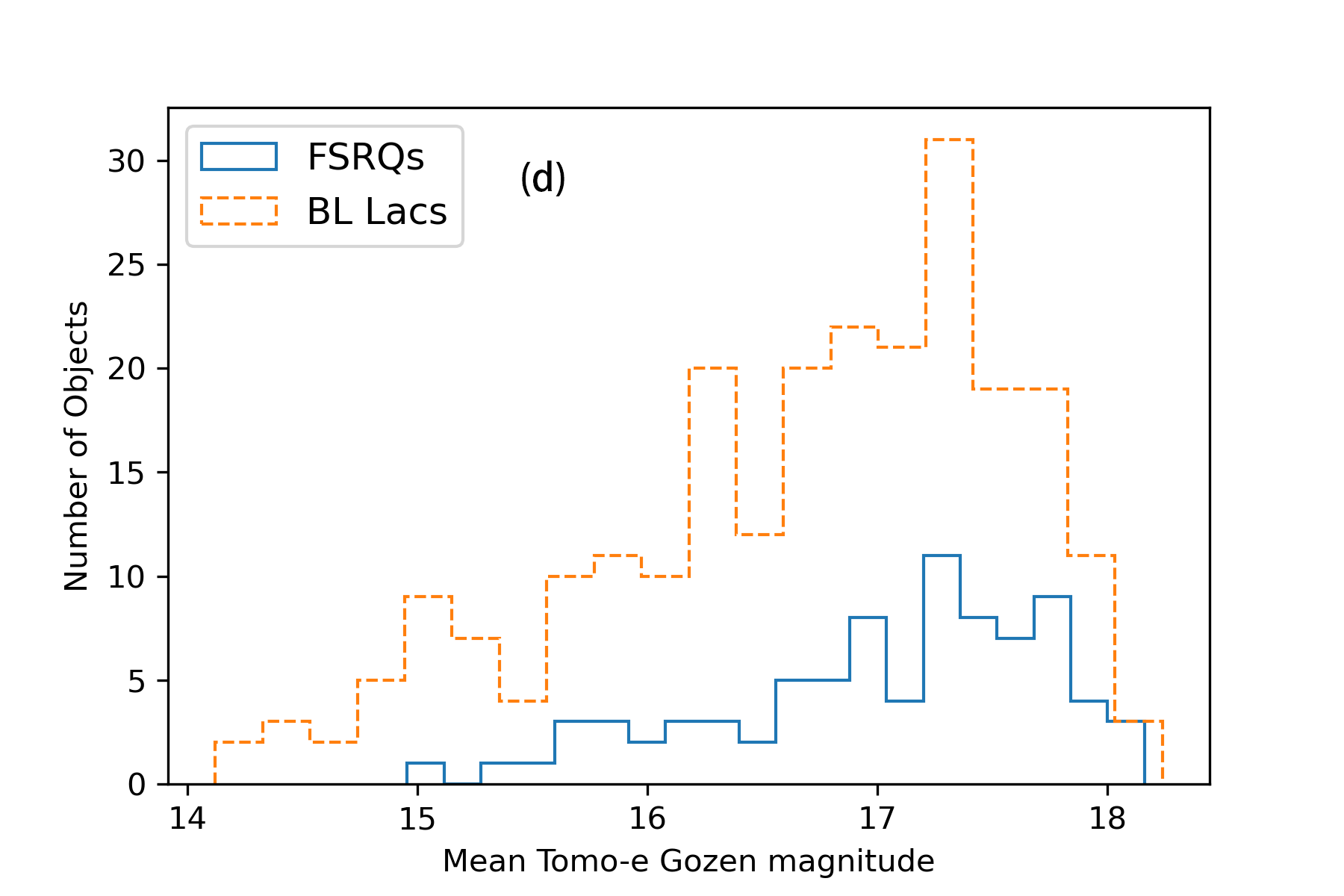}{0.5\textwidth}{(d)}
         }
\gridline{\fig{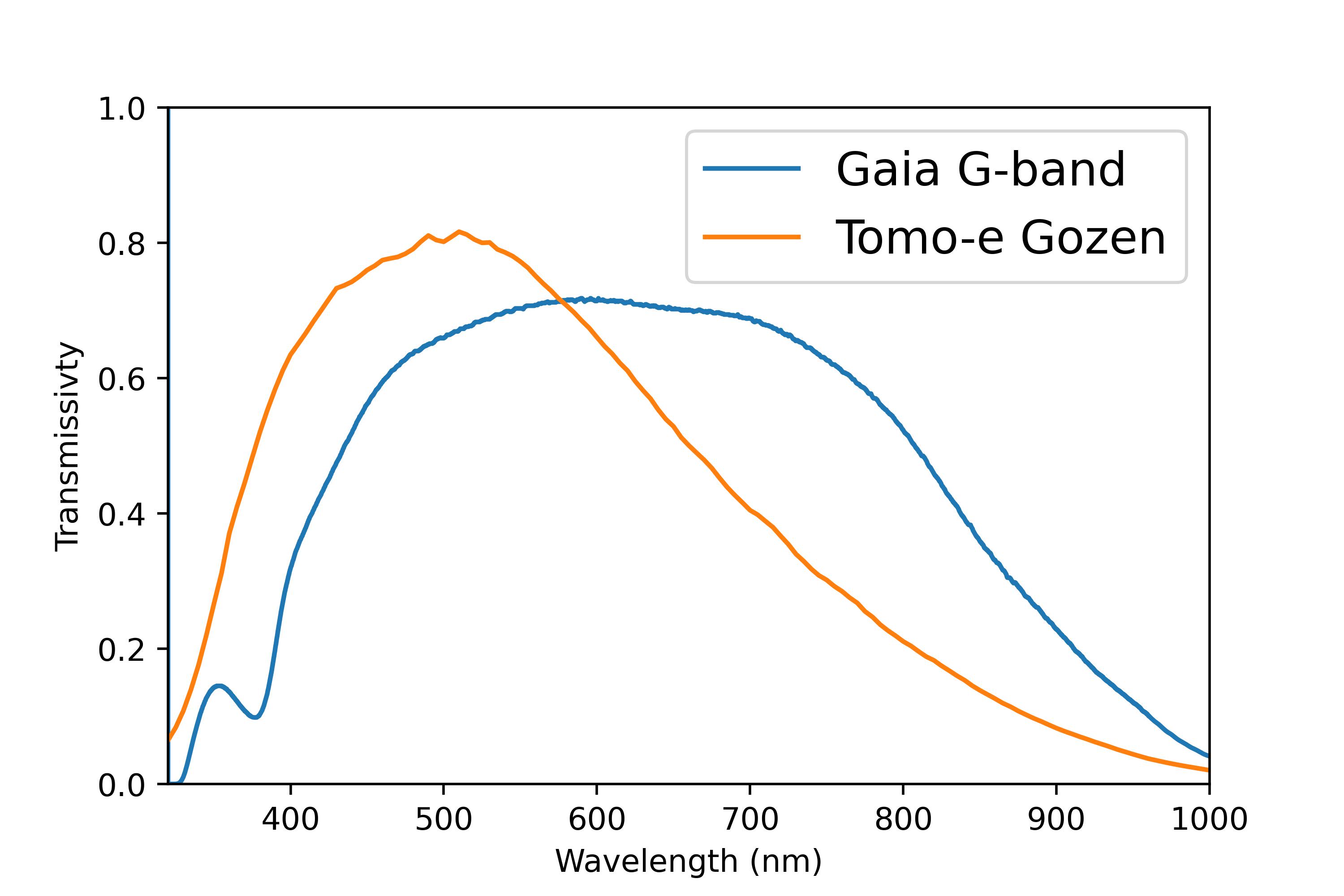}{0.5\textwidth}{(e)}
         }
\caption{(a): Photometric uncertainties of Tomo-e Gozen's photometry for SDSS standard stars. We obtained a total of 8,057,340 epochs of 283,605 SDSS standard stars between MJD 58726 and 59869. To determine the photometric uncertainty of Tomo-e Gozen at a given magnitude, we calculated the median of the standard deviation of repeated photometry for each star with the same r-band bin (0.01 mag) in the SDSS catalog; (b): The relationship between number of measurements and number of objects. In this study we only used blazars with more than 5 epochs; (c): Number of measurements pairs with time lag $\tau$ (day) of FSRQs and BL Lacs. (Due to the difference in redshift between BL Lacs and FSRQs, their $\tau$  distributions in the rest frame are different.); (d): Mean Tomo-e Gozen magnitude histogram of FSRQs and BL Lacs. (e): Bandpass of Tomo-e Gozen transparent window and Gaia G-band \citep{gaia2018gaia}.}
\label{fig:41}
\end{figure}

The Tomo-e Gozen Camera \citep{sako2018tomo} is a wide-field camera which
deploys 84 CMOS sensors and covers about 30$\%$ of the entire focal plane (9
degree in diameter) of the 1.05-m Kiso Schmidt telescope
\citep{takase1977105}.The camera has a peak photoelectric conversion efficiency
of 0.72 at a wavelength of 500 nm; the efficiency falls to 0.36 at wavelengths of 380 and 710
nm.(Figure \ref{fig:41}(e)). Since the CMOS sensor can read out the signal
pixel by pixel, the read out speed is fast ---
much more rapid than that of traditional astronomical CCDs.
We typically operate the camera in a high-speed mode,
reading out all 84 sensors at a rate of 2 frames per second.

In the 
Tomo-e Gozen project, most of the observation time is used for Northern Sky
Transient Survey (Tominaga et al. in preparation). The integration time 
for a single image is 0.5 seconds,
but we combine frames in groups of 12 (or 18 since August 14, 2020):
each frame is stacked with 11 (or
17) other images in its group 
after we have removed the maximum value at each pixel position 
to avoid cosmic rays.  
The effective exposure time for our measurements is therefore
6 seconds.
In our all-sky survey mode, the area covered each night 
is about 12,000 square degrees, and a sub-region with low atmospheric
extinction (above 40 degrees in elevation) is observed 2-3 times during the
night\footnote{Since December 5, 2019, we sometimes
operate in a high-cadence survey mode, covering  
$3000\,\mathrm{deg}^{2}$ more than 10 times every night at intervals
shorter than 1 hour,
rather than the all-sky survey mode.}
By making multiple observations of targets each night in this
way, we are exploring short-timescale transients such as supernovae and blazars.

To ensure the accuracy of photometric measurements in the Northern Sky
Transient Survey, we adopted SDSS standard stars given by
\citep{ivezic2007sloan} and observed with the Tomo-e Gozen camera. Following the
standard data reduction procedures such as dark and bias subtraction and
flat-field correction, we performed aperture photometry using an
aperture diameter of 16.6 arcseconds, 
compared to a typical Full Width at Half Maximum (FWHM) of 7
arcseconds. The magnitude zero point 
in our unfiltered images was calculated using G-band data from the
Gaia satellite DR2 \citep{gaia2018gaia}. We evaluated the photometric
uncertainty of the survey by determining the standard deviation of repeated
photometry across all observations for stars measured at least 10 times,
in the following manner.
We initially select all objects that are suitable for determining
the magnitude zero point (zeromag), based on various factors, such as 
their magnitude, parallax, and the
absence of other surrounding objects in the Pan-STARRS1 \citep{kaiser2002pan}
and Gaia catalogs. Subsequently, we
calculate the zeromag for each object in a frame and then use 
the median value as the overall zeromag of the frame. 
During this process, the standard deviation in
zeromag calculated for different objects is approximately 0.03 mag. The
zeromag error along with other errors (such as flat error, etc.) results in a
systematic error of 0.01 to 0.02 mag, and the overall photometric uncertainty
of the Northern Sky Transient Survey is presented in Figure \ref{fig:41}(a).

We chose as our initial source for
candidate blazars the Fourth LAT AGN Catalog (4LAC), derived from the first
eight years of data collected by the Fermi Gamma-ray Space Telescope within the
energy range from 50 MeV to 1 TeV \citep{ajello2020fourth}. 
The observational footprint of the Tomo-e Gozen
camera covers nearly 60$\%$ of the 4FGL survey's area, and its detection
capability down to mag 18.5 is able to identify 44.6$\%$ of the
blazar sources listed in the 4LAC.  Figure \ref{fig:41}(b) shows the number of
blazars observed and the histogram of the number of detections of each blazar
observed by the Northern Sky Transient Survey from 08-31-2019 to 11-04-2022.
Figure \ref{fig:41}(c) depicts a histogram of the time lag $\tau$ (days) in the
rest frame between two observations of the same blazar. The sparsity of
observational data in the $\tau \sim 100-250$ range can be attributed to the
seasonal visibility of objects, the climate of the Kiso Observatory, and
instrument adjustments of Tomo-e Gozen. Similarly, the scarcity of
observational data in the $\tau> 400$ range can be attributed to the limitation
of the survey period. Figure \ref{fig:41}(d), which displays a histogram of the
mean Tomo-e Gozen magnitude in repeated photometry for BL Lac and FSRQ blazars.
The distribution of magnitudes for BL Lac blazars in this study ranges from
magnitude 14-18, with a peak value of approximately magnitude 17.3. Similarly,
FSRQ blazars have magnitudes distributed in the range of magnitude 15-18, with
a peak value of approximately magnitude 17.3.

\clearpage

\begin{figure}[ht]
    \centering
    \includegraphics[scale=0.9]{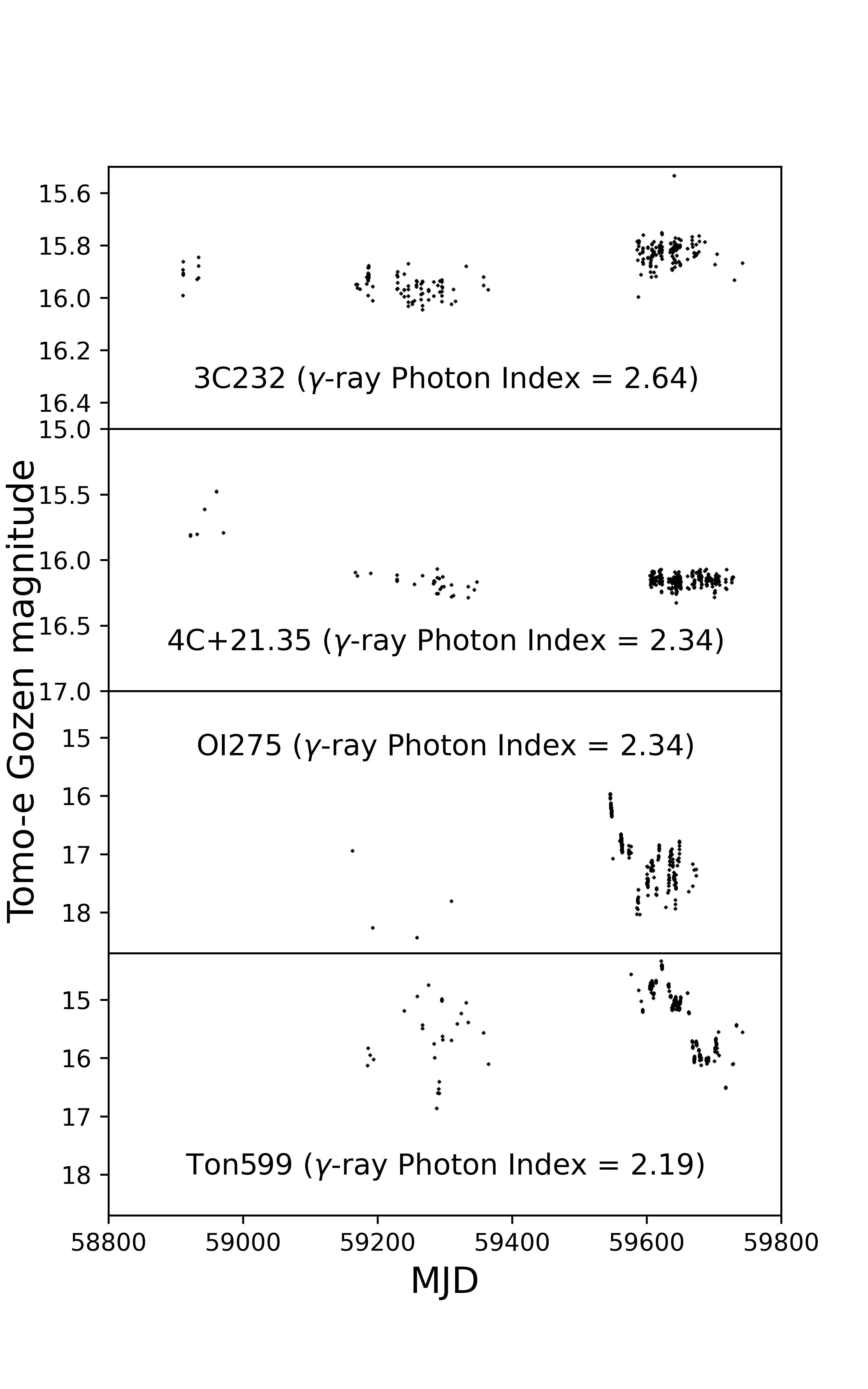}
    \caption{Light curves of well observed 4 FSRQs}
    \label{fig:fsrqlc}
\end{figure}

To facilitate the visual understanding of our analyzed dataset, we display
example light curves of four well-observed FSRQ objects in Figure
\ref{fig:fsrqlc}.
\clearpage
\subsection{Sample selection $\&$ variability measurement}
\label{sec:Data selection}

The luminosity relationship between blazars and their host galaxies has a
significant impact on our study. When the luminosity of the blazar is not
greater than that of the host galaxy, the continuous spectrum emitted by the
galaxy can overwhelm the emission lines from the AGN core of the blazars. This
phenomenon can also lead to an increase in the radiation photon index from the
jet, resulting in a redder spectrum. Consequently, this overlap in spectral
features can lead to the misclassification of certain FSRQs as members of the
BL Lac class, thereby introducing ambiguity and potential errors.

The assessment of variability of blazars is also complicated by the presence of
their host galaxies. First, the optical emission from the host galaxy merges
with that from the blazar, making it challenging to isolate the blazar's
variability. This is also particularly notable when the blazar's luminosity is
not significantly higher than that of the host galaxy, leading to potential
underestimation of the blazar's relative variability. Second, since we used a
fixed photometric aperture size (16.6 arcseconds in diameter), variations in
seeing can cause variations of the galaxy flux within the aperture, possibly
resulting in an overestimation of the blazar's absolute variability.

In order to minimize the potential for errors in
our analysis, we attempt to select objects in which
the luminosity contributed by the central engines (blazars) in our
sample will be greater than the luminosity contributed by their host
galaxies. A previous study \citep{urry2000hubble} employing the Hubble Space
Telescope to examine the host galaxies of BL Lac objects, spanning redshifts
from 0.1 to 1.3, revealed an absolute magnitude for these galaxies of $M_B =
-21.4 \pm 0.6$ AB mag (after adjustments for bandpass, Hubble constant, and
conversion from Vega to AB system). Hence the bright end of blazar host
galaxies within a 2$\sigma$ range is $M_B = -22.6$ AB mag. This finding is
consistent with bright end of the luminosity function of galaxies, as reported
by the SDSS survey \citep{blanton2003galaxy}, which is around -22.8 AB mag (in
the g-band). Consequently, we will exclude candidates whose optical luminosity
$L_O$ (including blazar and host galaxy) is lower than the K-corrected optical
luminosities of any hypothetical host galaxy type (including elliptical, Sc,
Sa, and starburst galaxies) corresponding to an absolute magnitude of $M_B=$
-23 AB magnitudes. 
We present in Figure \ref{fig:ol}
the K-corrected optical luminosity,$L_g$,
of different types of
hypothetical galaxies 
as a reference for galaxy
evolution at various redshifts;
we define $L_g$ as

\begin{equation}
    L_g=\frac{L_g^{\prime}}{k_g(z)}
\end{equation}
where $L_g^{\prime}$ is the optical luminosity of galaxies with absolute
magnitude $M_B=$-23 AB mag before K-correction and $k_g(z)$ is the K-correction
factor for galaxies with
\begin{equation}
    k_g(z)= (1+z)\frac{\int_{\nu_{1g}}^{\nu_{2g}} \frac{f_\nu(\nu)}{\nu} R_g(\nu)d\nu}{\int_{(1+z)\nu_{1B}}^{(1+z)\nu_{2B}} \frac{f_\nu(\nu)}{\nu} R_B(\nu)d\nu}
\end{equation}

where $\nu_{1g}$ and $\nu_{2g}$ are the starting and ending frequencies,
respectively, for the Gaia G-band, and $R_g(\nu)$ represents its
quantum efficiency. Similarly, $\nu_{1B}$ and $\nu_{2B}$ are designated as the
initial and terminal frequencies for the Gaia B-band, with $R_B(\nu)$ denoting its
quantum efficiency. $f_\nu(\nu)$ is the spectral density of flux of each galaxy
template provided by \cite{kinney1996template}.

\clearpage
\subsection{Optical luminosity of Blazers}
\label{sec:Optical luminosity of Blazers}

The optical luminosity of Blazars without K correction, $L_O^\prime$, was
calculated using redshift data from the 4LAC database,
average magnitude data from the Tomo-e Gozen survey. 
and the $astropy.cosmology.luminosity\_distance$ method
\citep{robitaille2013astropy}.
Power-law fitting was performed on the flux of each object in the
Pan-STARRS g, r, i bands \citep{kaiser2002pan} to determine its photon index
$\Gamma_O$
in the optical band, which was then used to perform K-correction 
and yield the optical
luminosity $L_O$ in the following manner:

\begin{equation}
    L_O=\frac{L_O^\prime}{k(z)}
\end{equation}
where
\begin{equation}
    k(z)= (1+z)\frac{\int_{\nu_{1t}}^{\nu_{2t}} \nu^{-\Gamma_O} R_t(\nu)d\nu}{\int_{(1+z)\nu_{1t}}^{(1+z)\nu_{2t}} \nu^{-\Gamma_O} R_t(\nu)d\nu}
\end{equation}
Here, $\nu_{1t}$ and $\nu_{2t}$ are the starting and ending frequencies of
the Tomo-e Gozen band and $R_t(\nu)$ represents Tomo-e Gozen's quantum efficiency
(see Figure \ref{fig:41}(e)).

In order to assess the variability of blazars, we first calculated the standard
deviation in repeated photometry (light curve). To obtain the intrinsic
standard deviation $\sigma$ of blazars, we subtracted the photometric
uncertainty, which is shown in Figure \ref{fig:41}(a), from this initial
standard deviation $\sigma'$. 
\begin{equation}
    \sigma=\sqrt{\sigma'^2-magerr(mag)^2}
\end{equation}
\begin{figure}[!ht]
\centering
\gridline{\fig{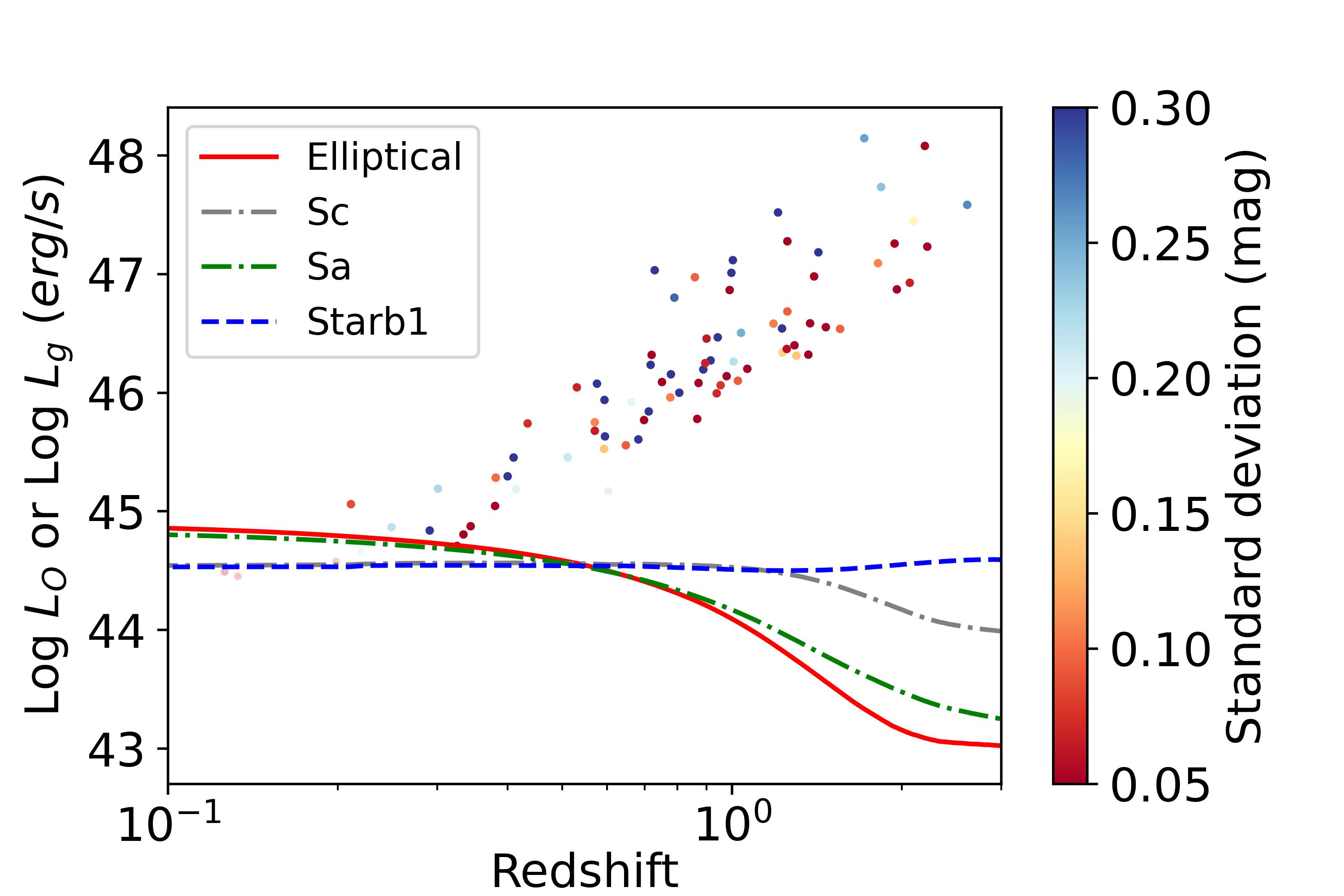}{0.5\textwidth}{FSRQs}
         \fig{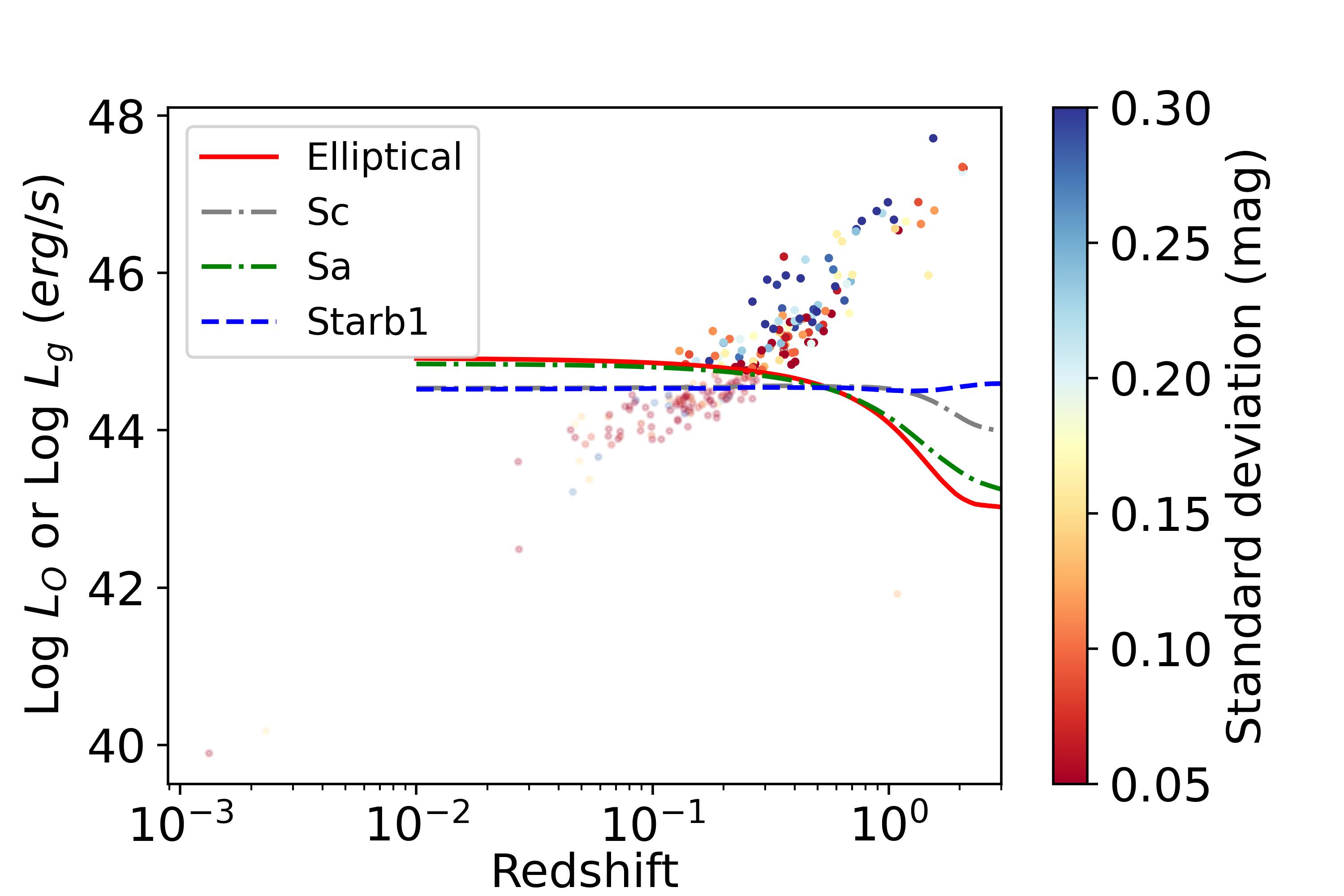}{0.5\textwidth}{BL Lacs}
         }
\caption{The distribution of K-corrected optical luminosity $L_O$, plotted against redshift for FSRQs and BL Lacs. The color of each symbol corresponds to the standard deviation of the repeated photometry (light curve). We have excluded blazars with $L_O$ below any of the $L_g$ lines, which represent the K-corrected optical luminosity of differents types of hypothetical galaxies with absolute magnitude of $M_B=$-23 AB mag. The symbols of the excluded blazers were blurred.}
\label{fig:ol}
\end{figure}

A total of 241 BL Lacs and 83 FSRQs were included in the original sample. To
ensure accurate observations of optical luminosity and avoid interference from
the stellar component of host galaxies, our study excluded blazars located below any of the lines
in Figure \ref{fig:ol} and made the final sample of 120 BL Lacs and 78 FSRQs
for more detailed investigation.

\clearpage
\section{Result \& Discussion }
\label{section:Variabilit Analysis}

\subsection{Standard deviation of repeated photometry}\label{sec: STD}

\begin{figure}[h]
    \centering
    \includegraphics[scale=1.2]{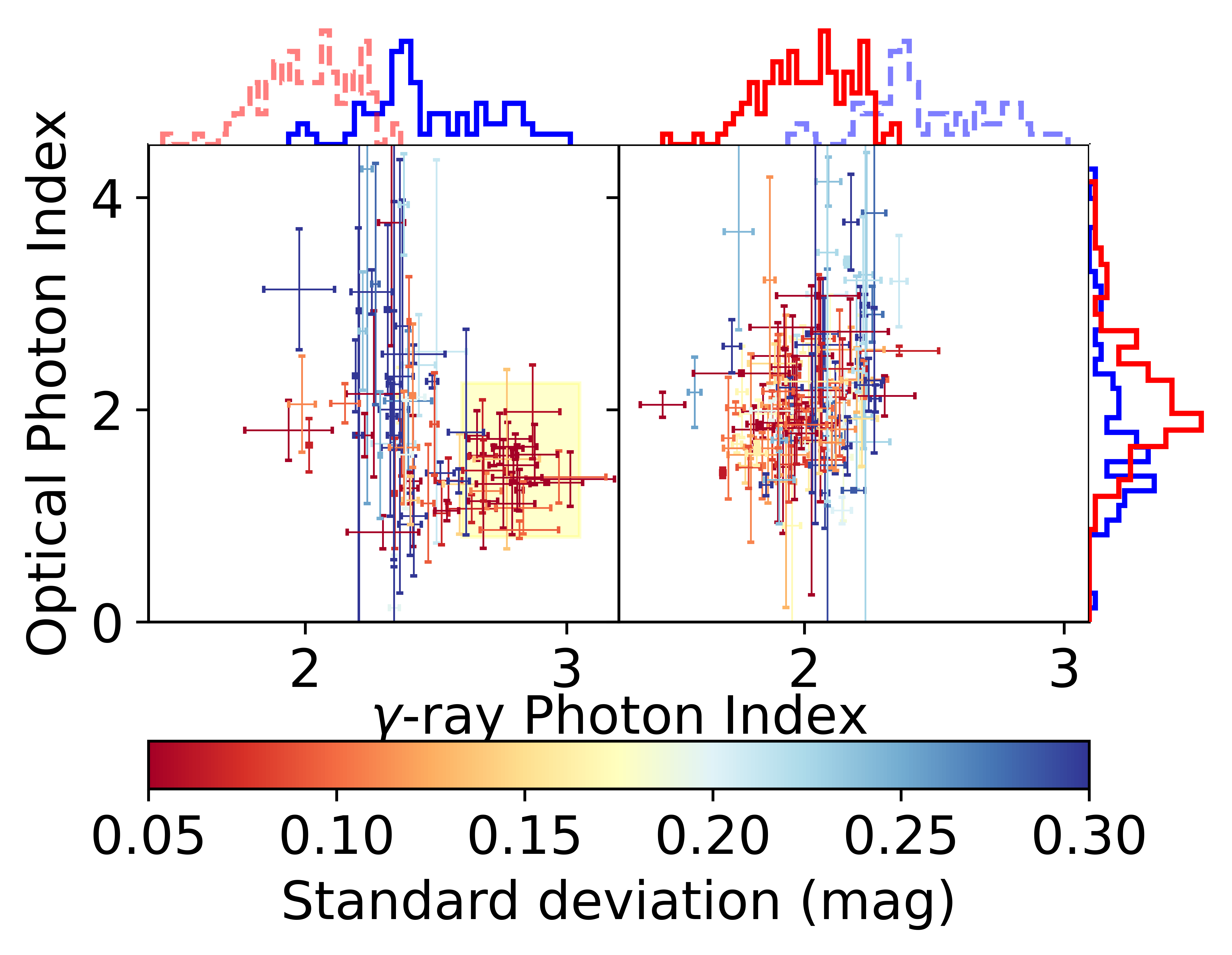}
    \caption{The relationship between the $\gamma$-ray photon indices (x-axis) and the optical photon indices (y-axis) is depicted for FSRQs (left panel) and BL Lacs (right panel). The top inset shows the histogram of the $\gamma$-ray photon indices, and the right inset displays the optical photon indices, with FSRQs indicated in blue and BL Lacs in red. The color intensity of each symbol reflects the standard deviation of the optical repeated photometry (light curve) for the respective sources. }
    \label{fig:pi}
\end{figure}

The distribution of photon indices of $\gamma$-ray\footnote{The
$\gamma$-ray photon index $\Gamma_\gamma$ taken from 4LAC is evaluated by
fitting a sub-exponentially cutoff power law $dN/dE \propto E^{-\Gamma}
\exp(a(E_{0}^b-E^b))$, where $a$, $b$ and $E_{0}$ are fitting parameters.} and
optical wavelengths\footnote{The optical photon index $\Gamma_o$ is defined  as
$dN/dE \propto E^{-\Gamma}$ or $f_{\nu} \propto \nu^{1-\Gamma}$, where $N$ is
the number of photons of the Pan-STARRS g, r, i bands \citep{kaiser2002pan},
$E$ is the photon energy, and $f_{\nu}$ is the flux density at the frequency
$\nu$.}, and their relationship to the standard deviation of repeated
photometry, are shown in Figure \ref{fig:pi}.

The distribution of optical photon indices for BL Lacs and FSRQs, while
overlapping in range, demonstrates a significant difference with a KS-test
p-value of 2.6e-5, possibly indicating multiple emission mechanisms. The peak
of the optical photon index distribution for FSRQs is notably lower, which may
be attributed to the bluer color of their accretion disks. Conversely, the
$\gamma$-ray photon index distribution shows a pronounced difference between
the two classes, with FSRQs typically exhibiting higher values than BL Lacs, as
indicated by a KS-test p-value of 6.6e-31. This disparity is likely due to the
differences in the peak of the inverse Compton scattering between the two. The
photon index of BL Lacs at $\gamma$-ray and optical wavelengths is
positively correlated, whereas those of FSRQs display an anti-correlation. Furthermore,
FSRQs with a $\gamma$-ray photon index greater than $\sim$2.6 (highlighted by
the yellow region in the left panel of Figure \ref{fig:pi}) present a smaller
standard deviation in repeated photometry measurements compared to those with a
lower $\gamma$-ray photon index. Further discussion of this trend is presented
in section \ref{subsection:sf}.

\newpage

\begin{figure}[h]
    \centering
    \includegraphics[scale=0.85]{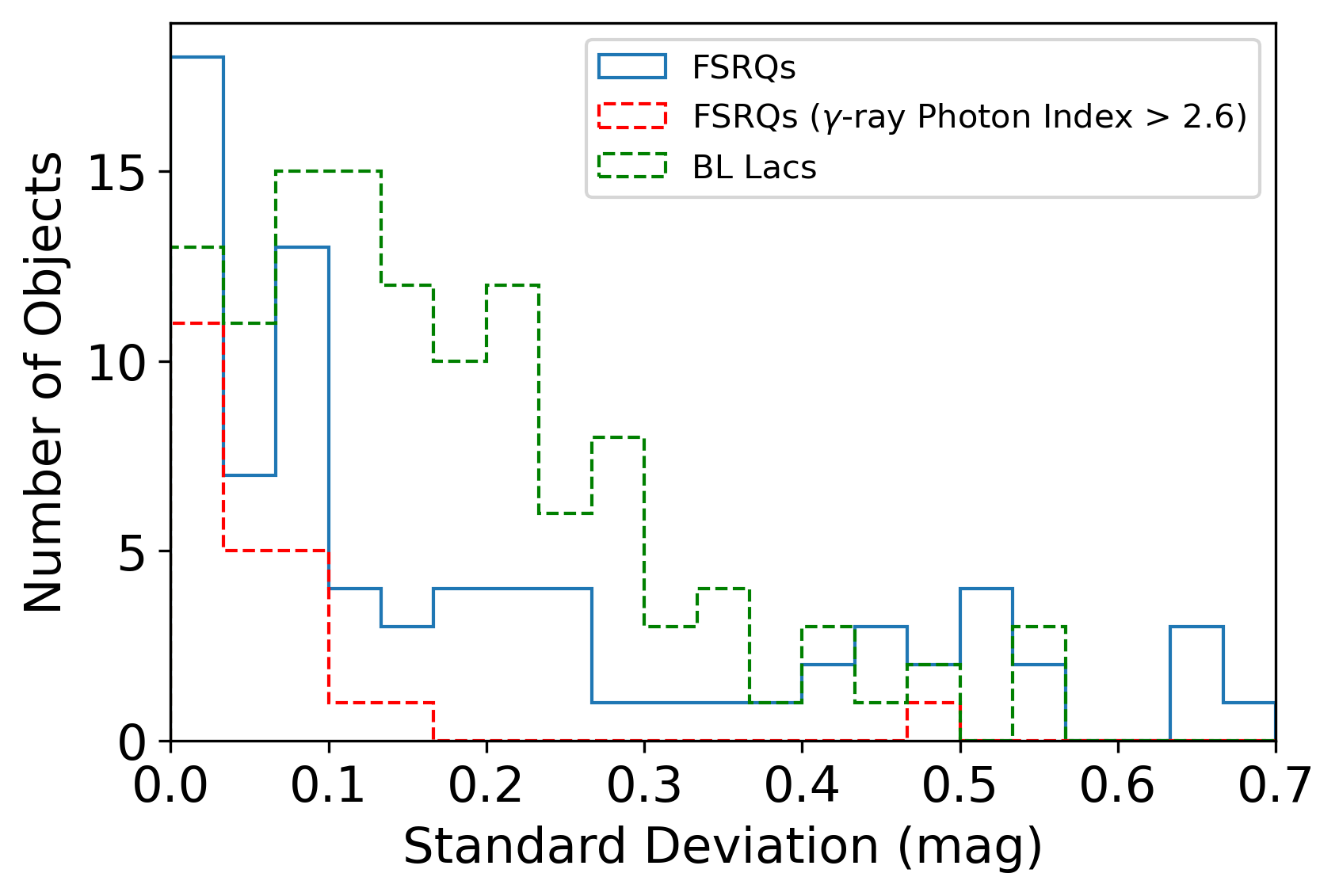}
    \caption{Distribution of standard deviation of repeated photometry of blazars which were observed more than 5 times. The blue line represents the overall FSRQs, the red line represents the FSRQs with high $\gamma$-ray photon indices (greater than $> 2.6$), and the green line represents BL Lacs.}
    \label{fig:big n std hist}
\end{figure}

In Figure \ref{fig:big n std hist}, we present a histogram that shows
the standard deviation observed in repeated photometric measurements of
blazars, specifically focusing on those observed more than five times. Our
analysis reveals a notable distinction in the variability patterns of BL Lacs
and FSRQs. The standard deviation for BL Lacs predominantly clusters within the
0.0 to 0.2 magnitude range. Conversely, the standard deviation for FSRQs
demonstrates a more uniform distribution across different magnitude bins.

Further scrutiny, employing the Kolmogorov-Smirnov (KS) test, accentuates the
disparity in the distribution patterns of FSRQs based on their gamma-ray photon
indices. FSRQs characterized by high gamma-ray photon indices (exceeding 2.6)
manifest a statistically significant divergence in their distribution when
compared to two groups: FSRQs with lower gamma-ray photon indices
(resulting in a p-value of 3.8E-6 ) and the general population of FSRQs
(yielding a p-value of 2.3E-3 ).

Figure \ref{fig:fsrqlc} provides detailed
illustrations of these relationships for several well-observed objects. 
The FSRQ 3C232, which has a high
$\gamma$-ray photon index, displays notable stability throughout our nearly
three-year observation period. Another largely quiescent object is 4C+21.35, 
which has a relatively
low $\gamma$-ray photon index; it exhibits slight fluctuations during the initial
stage of our observations (before MJD 59000) before becoming quiet in
subsequent observations. On the other hand, OI275 and Ton599, both of which
have relatively low $\gamma$-ray photon indices, display dramatic variations
during our observation period. The maximum optical luminosity of these two
objects can exceed the minimum optical luminosity by more than ten times.

\clearpage

\begin{figure}[!ht]
\centering
\gridline{
          \fig{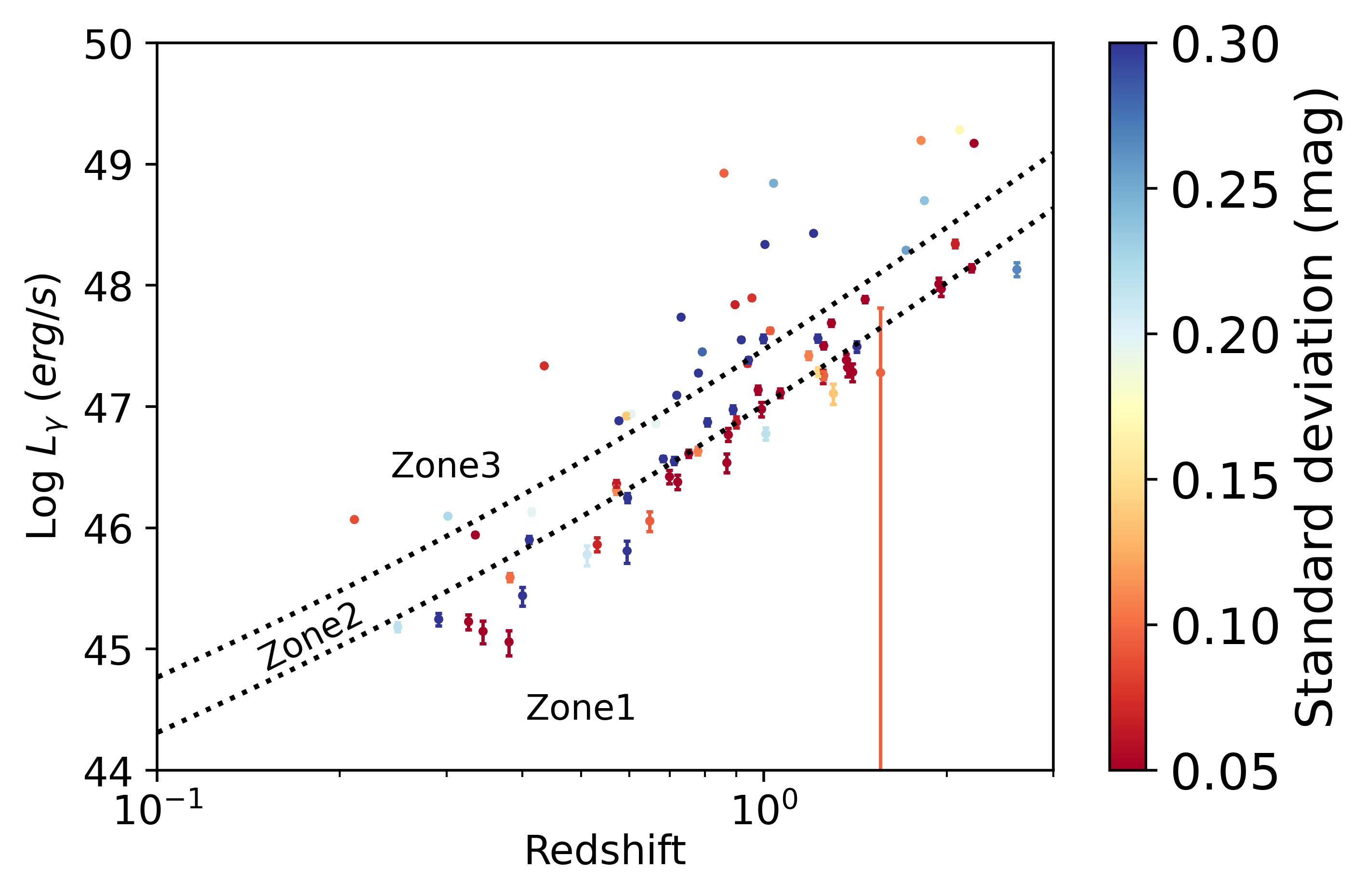}{0.5\textwidth}{FSRQs}
          \fig{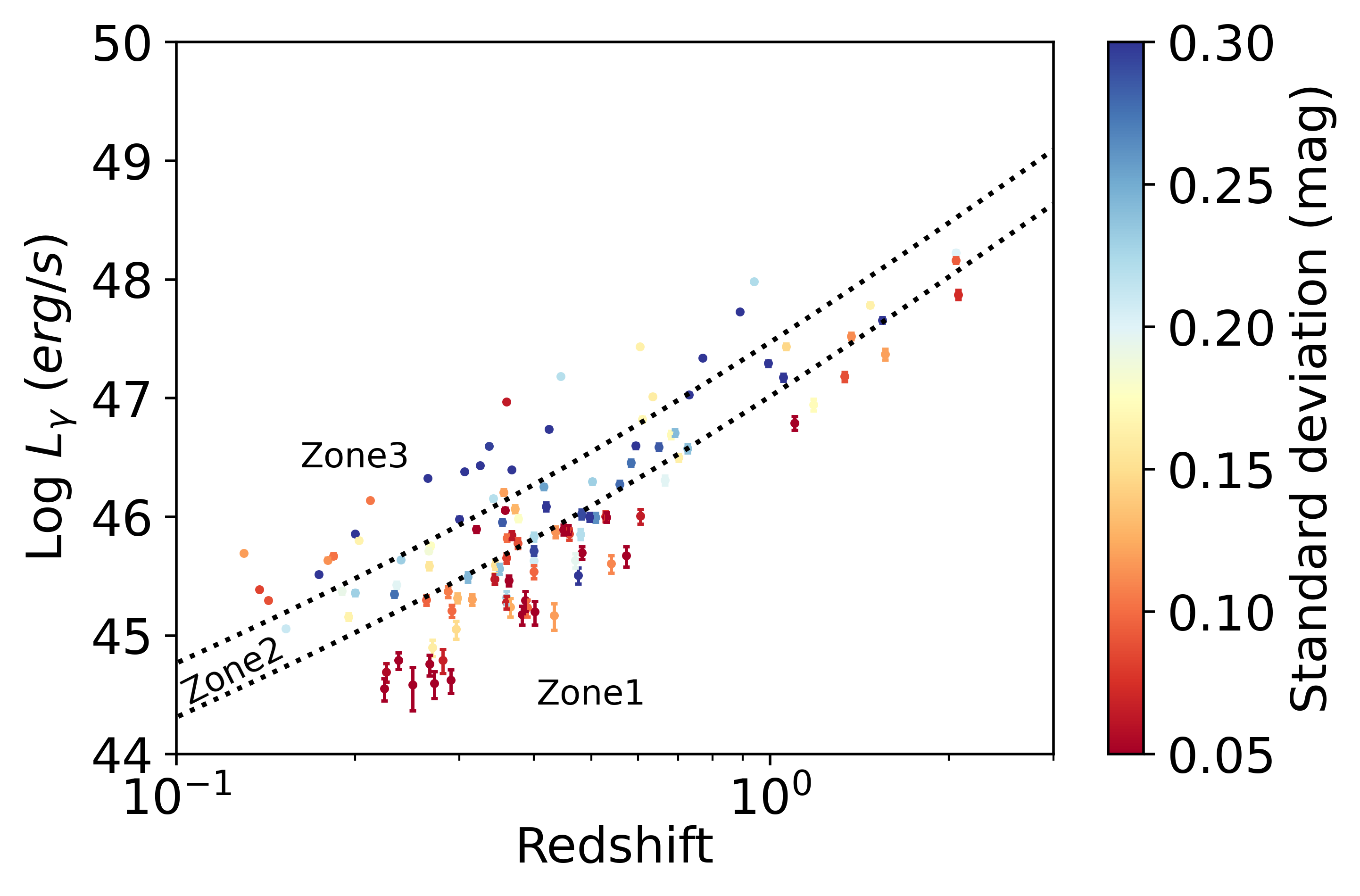}{0.5\textwidth}{Bl Lacs}
          }
\caption{The distribution of K-corrected $L_\gamma$ versus redshift of FSRQs and BL Lacs. The color of the symbol stands for the standard deviation of the repeated photometry (light curve). }
\label{fig:gl}
\end{figure}

Figure \ref{fig:gl} presents the relationship between redshift and
K-corrected\footnote{Similar to the K-correction of galaxies and optical band
but the spectral density of flux $f_\nu(\nu)$ will be $\nu^{1-\Gamma_\gamma}$
where $\Gamma_\gamma$ is the $\gamma$-ray photon indices from 4LAC catalog}
$\gamma$-ray luminosity ($L_\gamma$) for FSRQs and BL Lacs. Despite the
inevitable selection bias in the data due to sensitivity limitations and volume
effects, the limited dataset still reveals a noticeable trend: blazars with
lower optical variability (reddish symbols) tend to cluster in regions of lower $\gamma$-ray
luminosity ($L_\gamma$) across different cosmological epochs.

The analysis of blazars by \citep{ghisellini2017fermi} primarily utilized 
$L_\gamma$ as the distinguishing characteristic. While this approach offers a
straightforward method of categorization, it could overlook a critical aspect:
the potential diversity in physical properties of blazars at different
redshifts, despite having similar $L_\gamma$. This oversight is significant,
considering that the properties and surrounding environment of an AGN are
intricately linked to its redshift. Consequently, relying solely on $L_\gamma$
for classification, without considering the redshift, might lead to a skewed
understanding of blazar properties.

To quantitatively analyze the relationship between luminosity and redshift, we
introduce a new metric: the relative luminosity index ($R_L$). 
\begin{equation}
    R_L(L_\gamma,z)=log(\frac{L_\gamma}{4\pi k(z) D_L^2})+12
\end{equation}

where the term $k(z)$ represents the unified K-correction
coefficient\footnote{Similar to the K-correction of galaxies and optical band
but the spectral density of flux $f_\nu(\nu)$ will be $\nu^{1-\Gamma}$ where
$\Gamma$ is the average photon index specific to each blazar type; this equates
to 2.48 for FSRQs and 2.00 for BL Lacs.}
and $D_L$ denotes the luminosity
distance, which is also dependent on the redshift.

\begin{figure}[!ht]
\centering
\gridline{
          \fig{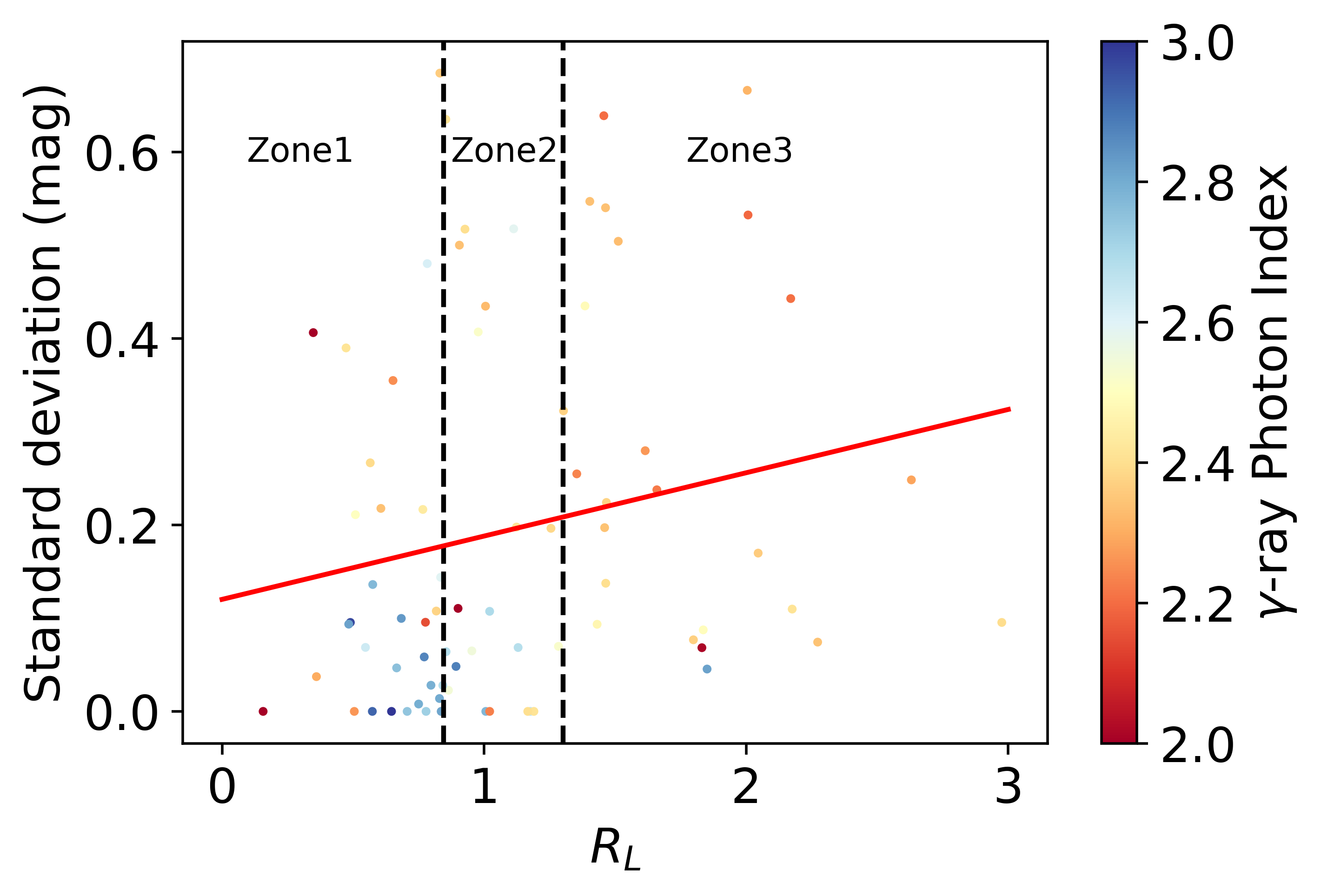}{0.5\textwidth}{FSRQs}
          \fig{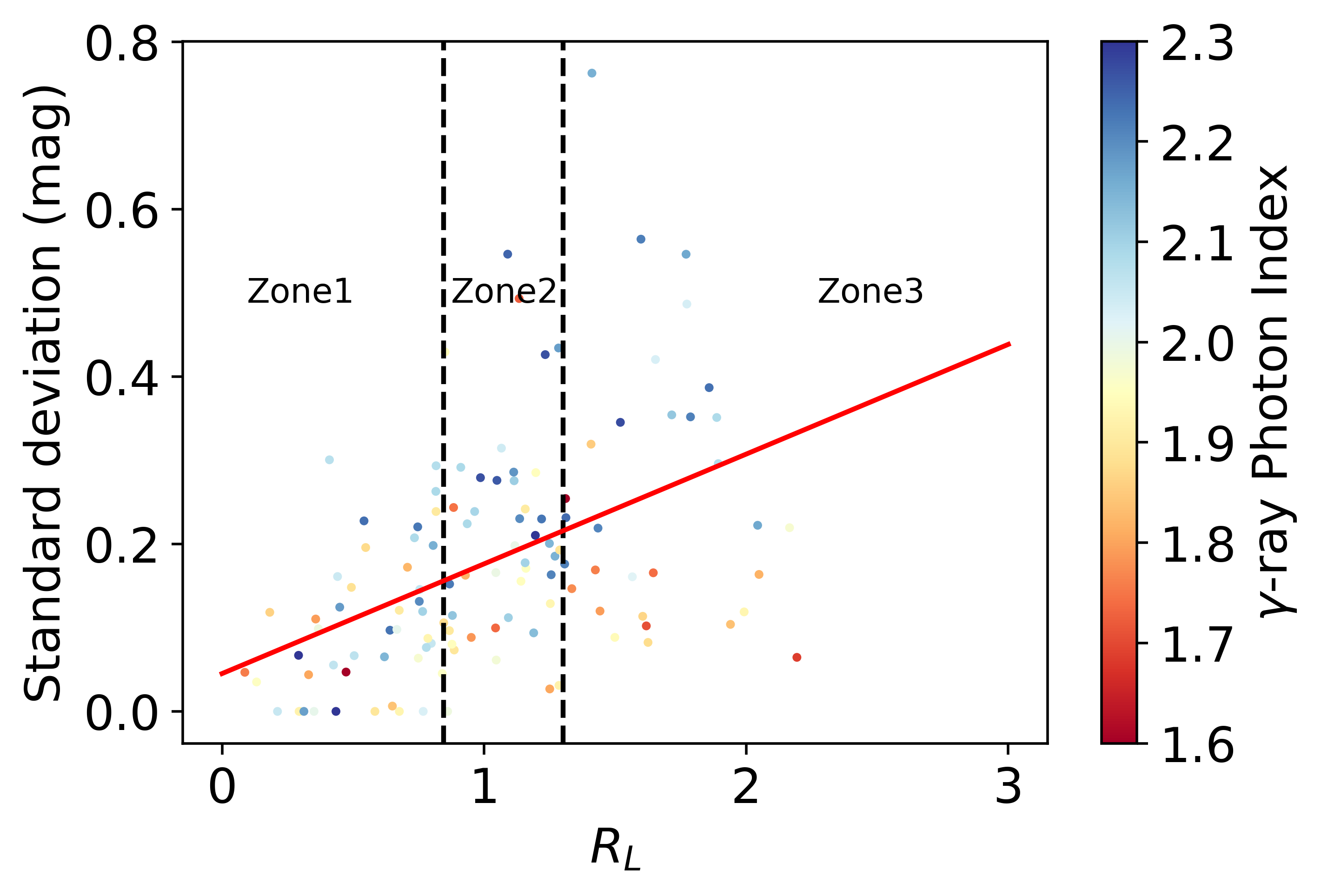}{0.5\textwidth}{Bl Lacs}
          }
\caption{The distribution of relative luminosity index ($R_L$) versus the optical variability of FSRQs and BL Lacs. The red solid line in the figure indicates the linear regression fit to the data. The color of the symbol stands for the $\gamma$-ray photon indices. }
\label{fig:zone}
\end{figure}

\clearpage
Figure \ref{fig:zone} delineates the correlation between the optical
variability of FSRQs and BL Lacs and their respective $R_L$ values. The red
solid line in the figure indicates a linear regression fit to the data. The
optical variability of both FSRQs and BL Lacs shows a positive correlation with
their relative luminosity index ($R_L$). For FSRQs, this correlation is weak
and not statistically significant, with a Pearson correlation coefficient of
r=0.19 and a confidence level of p=0.09. In contrast, for BL Lacs, the
correlation is moderate and statistically significant, with a Pearson
correlation coefficient of r=0.46 and a confidence level of p=1.6e-7.

This evidence for a difference between FSRQs and BL Lacs led us 
to look further into the relationship between the relative luminosity
index ($R_L$) and the dual-band spectral index. For a balanced and methodical
analysis, we categorized FSRQs into three distinct groups according to their
$R_L$ values, with the group boundaries set at log7 and log20. This
stratification was designed to ensure an approximately equal number of FSRQs in
each category. The same criteria for group classification were applied to BL
Lacs, thereby maintaining consistency and comparability in our analytical
approach. The FSRQs shown in Figure \ref{fig:fsrqlc} are located in different
zones. Specifically, 3C232 is located in zone 1, while OI275 is in zone 2. On
the other hand, 4C+21.35 and Ton599 are both in zone 3.

\clearpage
\begin{figure}[!ht]
\centering
\gridline{\fig{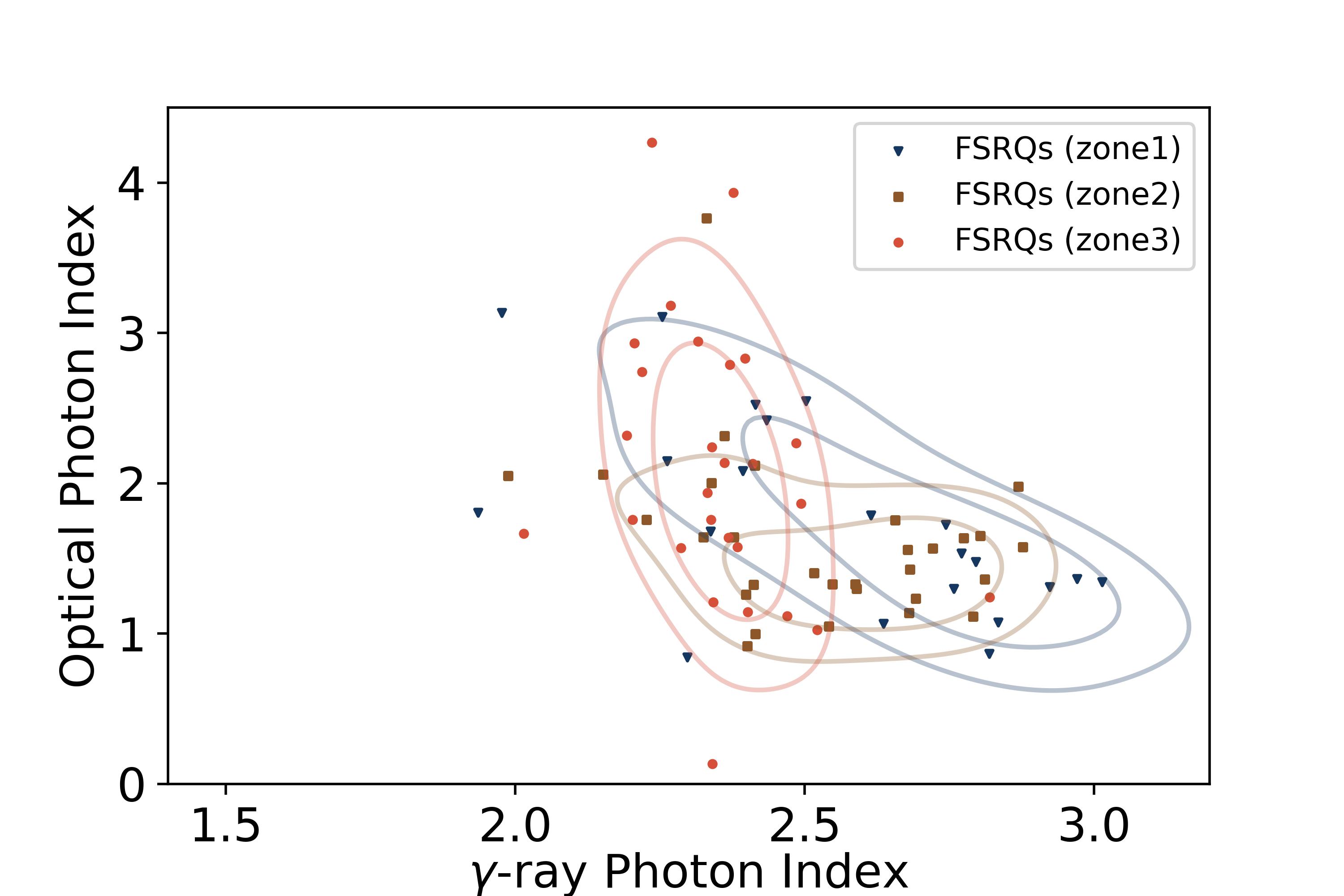}{0.5\textwidth}{FSRQs}
          \fig{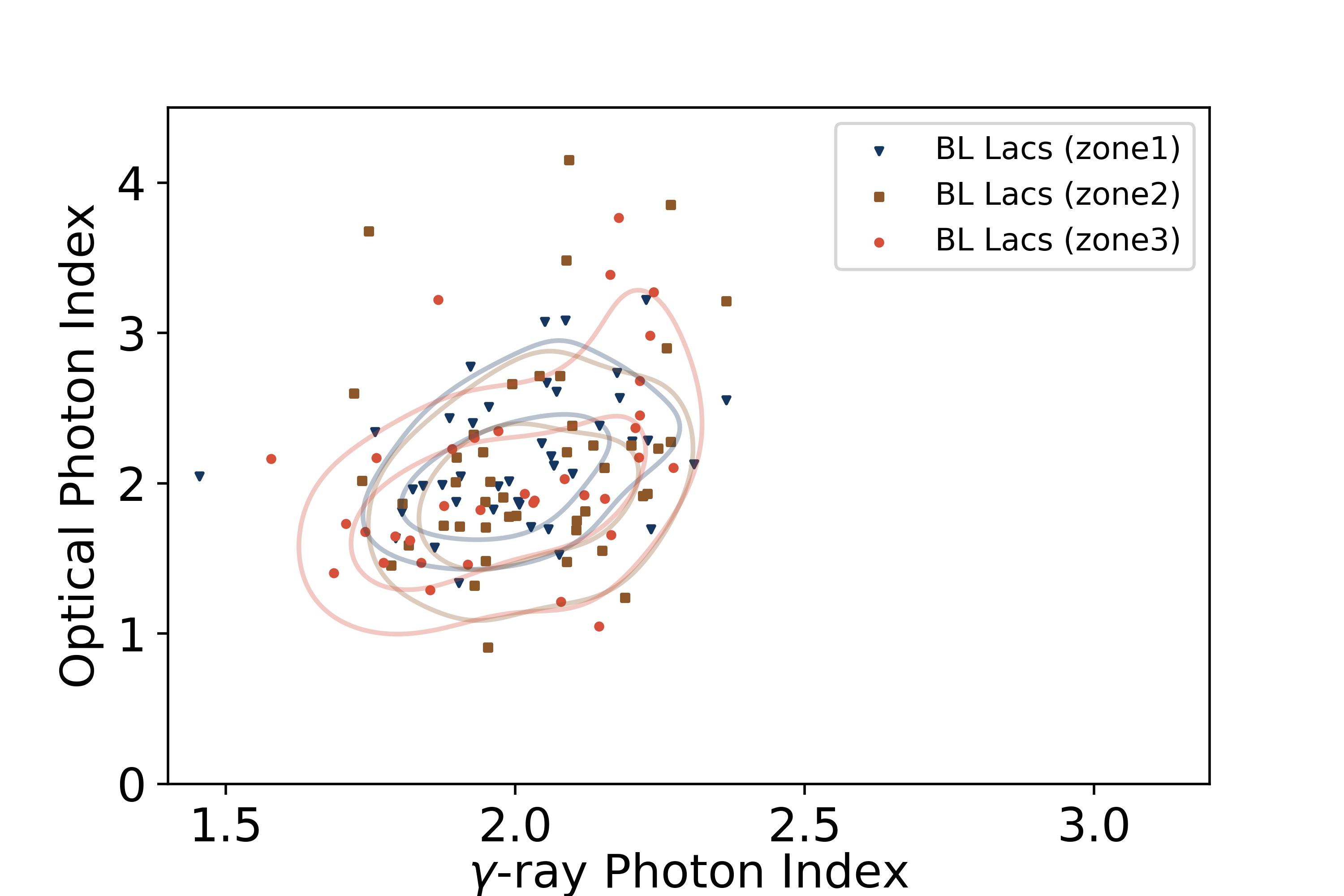}{0.5\textwidth}{BL Lacs}
          }
\caption{The distribution of the $\gamma$-ray photon index and the optical photon index of FSRQs (left) and BL Lacs (right) for each zone. (The kernel density estimate plot was generated using the sns.kdeplot function from the Seaborn data visualization library (version 0.12) in Python.)}
\label{fig:plzone}
\end{figure}

Figure \ref{fig:plzone} illustrates the distribution of $\gamma$-ray photon
index and optical photon index for FSRQs and BL Lacs in each zone. Zone 1
FSRQs dominate the high $\gamma$-ray photon index group, suggesting that these
FSRQs tend to have lower $L_\gamma$ compared to other FSRQs. Conversely, the
low $\gamma$-ray photon index group does not appear to be dominated by any
specific zones of FSRQ, as all three zones exhibit similar tendencies. The
optical photon index of the high $\gamma$-ray photon index FSRQs is
concentrated around 1.5, which is the typical color of a quasar accretion disk
\citep{kishimoto2008characteristic}. This suggests that FSRQs dominated by
accretion disks differ significantly from those dominated by jets in terms of
optical variability and $\gamma$-ray photon indices. Notably, zone 2 FSRQs also
have optical photon indices concentrated around 1.5, despite their relatively
scattered $\gamma$-ray photon indices, while zone 3 FSRQs have a relatively
concentrated $\gamma$-ray photon indices ($\sim$ 2.3) but have relatively
scattered optical photon indices.

\clearpage
\subsection{Structure function}
\label{subsection:sf}
In addition to the standard deviation of repeated photometry, which does not
involve the rest frame time interval $\tau$, we also studied the structure
function, following the method by \citep{berk2004ensemble}. The structure
function can be defined as

\begin{equation}
\mathrm{SF}(\tau)=\sqrt{\left\langle[m(t)-m(t-\tau)]^{2}\right\rangle-\left\langle\sigma_{SF}^{2}\right\rangle}
\end{equation}
where $m(t)$ and $m(t-\tau)$ are magnitudes of two measurements of the same
object, separated by (rest frame) time $\tau$, 
and $\sigma_{SF}$ is uncertainty of the
measurements given by (\ref{equ:err}), where $magerr(t)$ represents the blue
line in Figure \ref{fig:41}(a).
\begin{equation}
\sigma_{SF}=\sqrt{magerr(t)^2+magerr(t-\tau)^2}
\label{equ:err}
\end{equation}
The optical variability of an object can be quantified using the structure
function, which depends on the rest frame time interval $\tau$. This function
provides insight into the temporal characteristics of the object's variability.
A change in the slope of the structure function 
when plotted again $\tau$ may signify a significant
timescale relevant to understanding the underlying mechanism of optical
luminosity variation. To construct the structure function for a single object, a
large number of observation epochs that are widely distributed in time are
required. However, due to limited data availability, 
we constructed structure
functions for a class of blazars as a group,
combining measurements from objects with similar properties.

\begin{figure}[!ht]
    \centering
    \includegraphics[width=14cm]{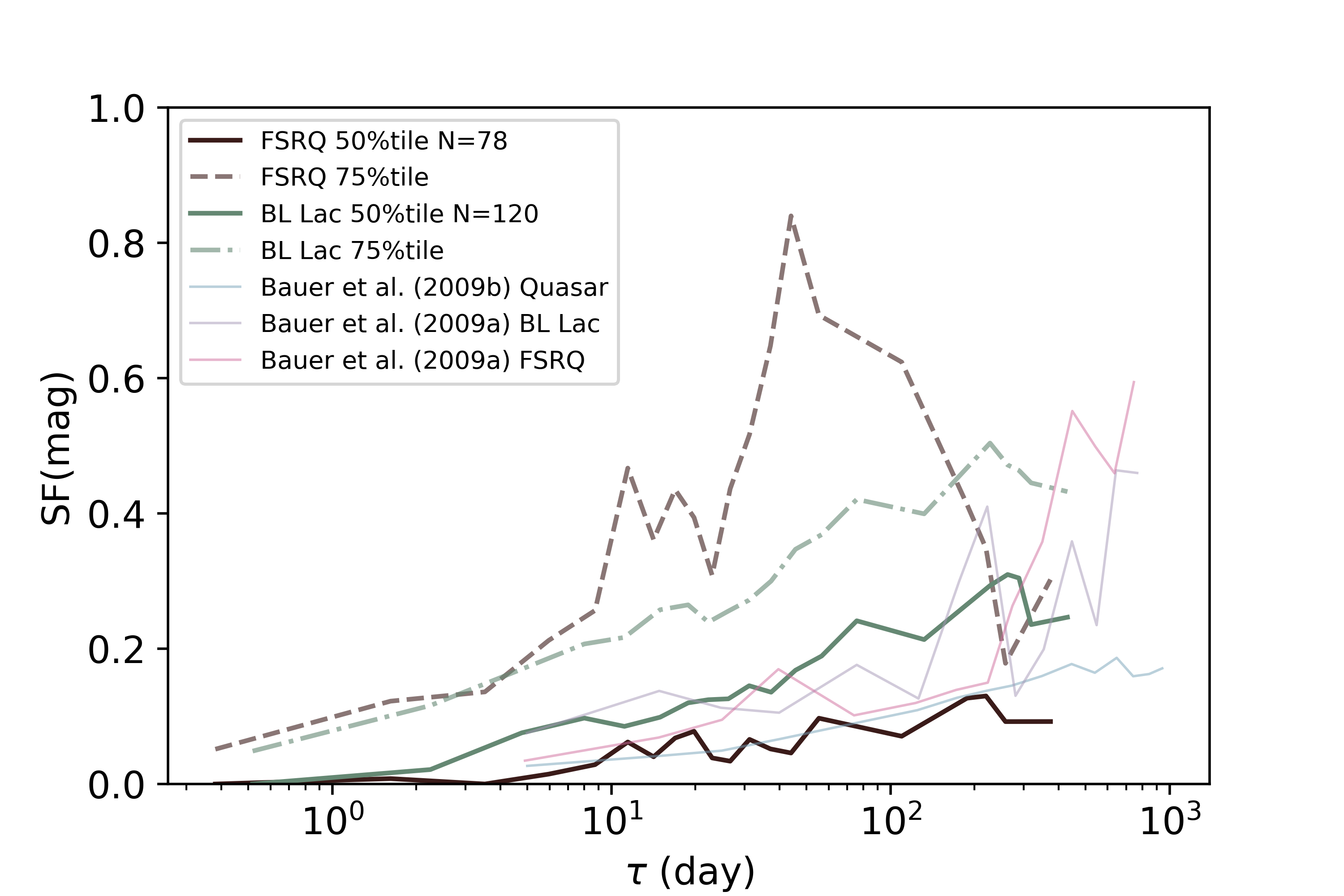}
    \caption{Structure function of FSRQs (brown; 10966 data pairs for each point) and BL Lacs (green; 34015 data pairs for each point), and previous results from \citep{EmmanoulopoulosBauer_2009} and \citep{bauer2009quasar}. The rest frame interval $\tau$ is in unit of day.}
    \label{fig:sf}
\end{figure}

Figure \ref{fig:sf} illustrates the structure functions of FSRQs and BL Lacs.
In the process of constructing the structure function for a specific class of
blazars, either BL Lacs or FSRQs, a comprehensive approach was adopted to
extract and analyze the observational data. The initial step involved the
extraction of all available data pairs 
(magnitudes of one object at two epochs) from the selected blazar class,
recording their respective rest frame time lag $\tau$ and structure function
values. This extensive data set was then combined into a unified set,
encompassing observations from various sources.

To facilitate a detailed analysis, the combined data pairs were organized in
ascending order based on their $\tau$ values. The ordered data set was then
equally divided into 20 groups to ensure a balanced representation across
varying lag intervals. For each of these groups, a representative data point
was generated to populate the structure function graph. The x-coordinate of
these data points, indicating the lag value, was calculated as the mean of all
lag values within the respective group. 
Within each group, we computed 
two y-coordinates:
one for the 50th percentile of the SF values,
the other for the 75th percentile of the SF values.

In the rest-time time interval $\tau$ 
spanning 10 to 100 days, it is evident that the
median (50th percentile) variability of Flat Spectrum Radio Quasars (FSRQs) is
lower than that of BL Lacertae objects (BL Lacs), while their 75th percentile
variability exceeds that of BL Lacs. This observation suggests that FSRQs
exhibit greater internal heterogeneity than BL Lacs. Despite this 
difference, it's important to note that both BL Lacs and FSRQs have a similar
smooth pattern in the intraday timescale. Additionally, contrary to the
findings of Emmanoulopoulos and Bauer (2009), no distinct trough in variability
was observed in the 70-250 day range. 
We attribute this discrepancy to the
expanded dataset now available and the refined criteria for sample selection in
our study.  

\clearpage

\begin{figure}[!ht]
\centering
\includegraphics[scale=1]{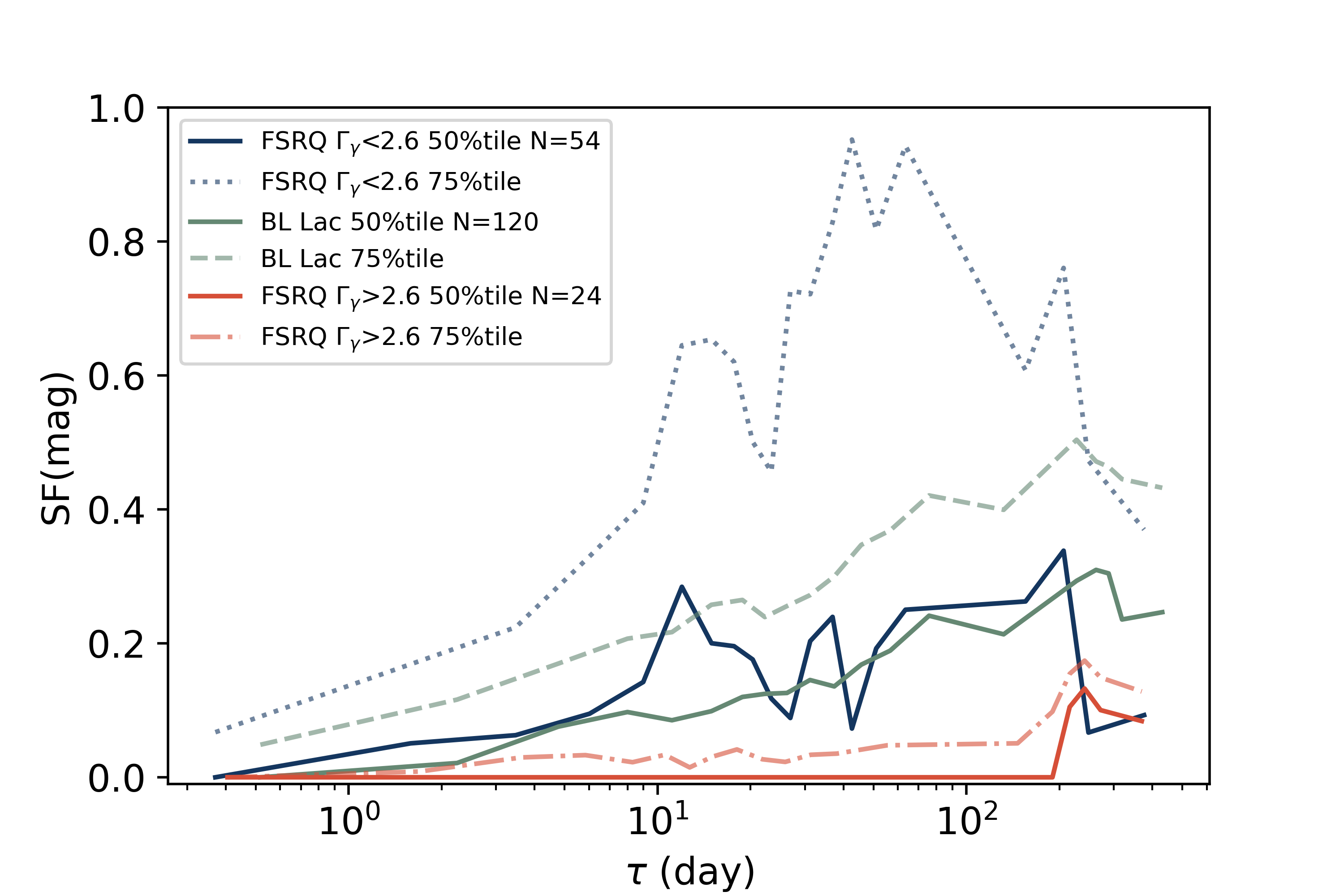}
\caption{Structure function of low $\gamma$-ray photon index FSRQs (blue; 7689 data pairs for each point), BL Lacs (green; 34015 data pairs for each point) and high $\gamma$-ray photon index FSRQs (red; 3278 data pairs for each point).}
\label{fig:sfpl}
\end{figure}

We also performed further analysis on FSRQs 
grouped by their $\gamma$-ray photon index,
as mentioned in Section \ref{sec: STD}; 
see Figure \ref{fig:sfpl}.

In this comparative analysis, the amplitude of the structure function for high
$\gamma$-ray photon index FSRQs stands out for its significantly lower value
when compared with both BL Lacs and the low $\gamma$-ray photon index FSRQs.
Notably, the structure function of high $\gamma$-ray photon index FSRQs shows
substantial variability only beyond a 200-day timescale. This observation
aligns with the hypothesis that the variability in these FSRQs resembles that
of accretion disks, which are characterized by timescales over 100 days
\citep{burke2021characteristic}. 
In contrast, for low $\gamma$-ray photon index
FSRQs, optical variability within a shorter span, particularly under 10 days,
is likely attributed to jet components (\citep{penston1970optical};
\citep{fan2005optical}). Furthermore, figure \ref{fig:sfpl} reveals that low
$\gamma$-ray photon index FSRQs (sometimes
thought to be dominated by an accretion
disk) exhibit more significant variability than the jet-dominated BL Lacs,
particularly within the 10- to 100-day range (KS-test p$<$1E-16). This finding
suggests a complex role of accretion disks and jets in these sources, pointing
towards the necessity of further spectroscopic observations to determine the
dominant factor driving these changes.

\clearpage
\section{Discussion} \label{Discussion}
Based on our analysis, we infer that the optical component of FSRQs
with high $\gamma$-ray photon index is dominated by the accretion disk. This
hypothesis is based on several observational results: their low optical
variability, their lower $L_\gamma$ compared with other FSRQs in the same
cosmic period, and their accretion-disk-like characteristic timescale. These
findings are consistent with the results of\quad\citep{shaw2012spectroscopy},
who reported a weak correlation between non-thermal dominance of FSRQs and low
photon indices in Fermi observations. Unfortunately, due to the limited sample
size, 
our study only separated FSRQs by $\gamma$-ray photon index into two
groups. Our result is different from the result of previous studies
\citep{ghisellini2017fermi} which only detected accretion disc components in
spectra (from ASI Astrophysical Data Center) of FSRQs at high redshift, while
FSRQs at low redshift rarely showed accretion disc components in the optical
band. Their result may be attributed to the accretion disc component becoming
more prominent at larger redshifts, thus at higher luminosities. However our
study provides the first confirmation of accretion disk-dominated FSRQs at
low redshifts through their low optical variability and 
long characteristic timescale.
It remains unclear whether there is a continuous relationship between the
$\gamma$-ray photon index and the dominance of the accretion disk, 
or if the high
accretion-disk dominance ratio only appears in the high $\gamma$-ray photon
index region.

\citep{ghisellini2016blazar} argued that the IR-optical band SED of FSRQs with
low $L_\gamma$ contains smaller accretion-disk components than other FSRQs, and
the flux at optical wavelengths is dominated by non-thermal radiation of
the jet, which differs from the results of this study. The main reason for this
difference may be our different blazar selection and grouping approach.
\citep{ghisellini2016blazar} grouped the FSRQs only according to their
$L_\gamma$, while in this study, we considered both the $L_\gamma$ and redshift
of the blazar simultaneously. The reason for this approach is that we believe
the nature of blazars with the same $L_\gamma$ is not precisely identical but
closely linked to their cosmic epoch.

It is important to acknowledge that the $\gamma$-ray photon index utilized in
our study are based on the Fermi/LAT eight-year integration, which represents an
average of the blazar $\gamma$-ray spectrum over an eight-year period.
Conversely, the optical spectrum index utilized in our study is based on the
result of short-term observations by Pan-STARRS1, indicating that our two photon
indices are not synchronized. Consequently, caution must be exercised in
interpreting the results of our study, particularly with regards to the
correlation between the $\gamma$-ray and optical photon index properties.

Also, according to the blazar sequence given by \citep{ghisellini2017fermi},
for FSRQs and BL Lacs at high redshift, the frequency observed by Tomo-e
Gozen (about $10^{14}$ Hz) is higher than their synchrotron radiation peak
$\nu_s$ (about $10^{12}$ Hz), while for BL Lacs at low redshift, the
frequency observed by Tomo-e Gozen is higher 
than their synchrotron radiation peak
$\nu_s$ (about $10^{16}$ Hz). This redshift dependence may lead to differences
in observed optical variability. However, considering that the BL Lacs in the
sample have roughly the same redshift distribution as the FSRQs, and the
standard deviation of repeated photometry of both FSRQs and BL Lacs did not
appear to be redshift-dependent, we conclude that this phenomenon does not
significantly affect our results. 

\clearpage
\section{Conclusion}\label{Summary}
We conducted a study of the optical variability of blazars 
selected from the 4LAC catalog,
using observations from the
Tomo-e Gozen Northern Sky Transient Survey, covering the period from MJD 58726
to 59887. 
Our high-speed sky survey allowed us to
observe 324 blazars, with an average of 50 epochs per blazar, representing a
significant increase compared to the previous study by
\citep{EmmanoulopoulosBauer_2009}, which had an average of only six epochs per
blazar. Furthermore, the catalog used in our study was updated, which provided
additional information not present in the previous study.

We focused exclusively on blazars with an optical luminosity $L_O$ higher than
that of hypothetical galaxies with absolute magnitudes of $M_B$=-23 (AB
magnitude). This criterion was applied 
in order to enhance the probability of
the correct spectroscopic classification of low-luminosity blazars and
accurately calculate their optical variability.

Our study shows that there is no significant difference in the optical
variability of blazars at different cosmological epochs, despite the orders of
magnitude difference in their luminosity. This finding indicates that the
optical variability of blazars is not directly dependent on their luminosity,
nor on their redshift. 
As we considered the redshift dependence and $L_\gamma$
dependence simultaneously in our analysis, we found that blazars with low
$L_\gamma$ at a given cosmic epoch tended to have lower optical variability.

The $\gamma$-ray photon index of FSRQs is typically higher than that of BL
Lacs. We found that the photon index of BL Lacs is positively correlated
at $\gamma$-ray and optical wavelengths, while the that of FSRQs 
is anti-correlated.
Also, FSRQs with high $\gamma$-ray photon index tend to have lower optical
variability, especially those with $\gamma$-ray photon index exceeding 2.6.
Moreover, the distribution 
of standard deviation in repeated photometry of FSRQs with $\gamma$-ray photon
index greater than 2.6 is significantly different from that of 
FSRQs with $\gamma$-ray photon index less than 2.6. The
difference between these two distributions is statistically significant with a
KS test P-value equal to 2.6e-6.

FSRQs with high $\gamma$-ray photon index (greater than 2.6) typically
exhibit low optical variability. From an evolutionary perspective, these FSRQs
are often found in regions of low $L_\gamma$ at a cosmic epoch, a pattern that
aligns with what is observed in disk-dominated blazars. While their
characteristic timescales are not immediately apparent from our observation
alone, further observations and analysis will, we hope provide more insight.

When examining the structure functions of these FSRQs in comparison to 
those of BL Lacs,
a clear distinction emerges: FSRQs generally show higher amplitudes in their
structure functions. Specifically, FSRQs with high $\gamma$-ray photon index
have significantly lower amplitudes compared to their low-index counterparts.
This finding is intriguing as these high-index FSRQs also display noticeable
variability, predominantly on timescales extending beyond 100 days. This
pattern of variability, emerging from the analysis of structure functions,
suggests a resemblance to the longer timescales typically associated with
accretion disks. In contrast, FSRQs with low $\gamma$-ray photon index
exhibit variability on much shorter timescales, often less than ten days,
likely indicative of jet activity. Moreover, these low $\gamma$-ray photon
index FSRQs demonstrate even greater variability than jet-dominated BL Lacs,
not only on timescales less than ten days but also particularly within the 10
to 100-day range  (KS-test p$<$1E-16), pointing towards an area 
worth 
further exploration to determine the underlying causes. 
Despite these
differences, it's important to note that the structure functions of both BL Lac
and FSRQs consistently exhibit smooth characteristics at the intraday scale.

We are initiating spectral studies of FSRQs, particularly focusing on emission
line equivalent widths, to decipher the roles of accretion disks and jets. In
parallel, we will continue our engagement with the Tomo-e Gozen Northern Sky
Transient Survey to enrich our dataset. This sustained effort is expected to
enhance our understanding of blazar luminosity variability and further
illuminate the complex mechanisms driving these FSRQs.

\begin{acknowledgments}
This work has been supported by the Japan Society for the Promotion of Science (JSPS) KAKENHI grants,  24K17097, 23H04894, 21H04491, 18H05223, 18H01272, 18H01261, 18K13599, 17H06363, 16H06341, 16H02158, 26247074, and 25103502. This work is supported in part by the Optical and Near-Infrared Astronomy Inter-University Cooperation Program (OISTER), the Ministry of Education, Culture, Sports, Science and Technology (MEXT) of Japan. We acknowledge the use of data from the Fermi Gamma-ray Space Telescope/Large Area Telescope (Fermi/LAT) the Sloan Digital Sky Survey (SDSS), the Panoramic Survey Telescope and Rapid Response System (Pan-STARRS), and the Gaia mission. Fermi/LAT is a collaboration between NASA and the Department of Energy in the United States, and is supported by institutions in France, Italy, Japan, and Sweden. This work was supported in part by Japan Foundation for Promotion of Astronomy. SDSS is managed by the Astrophysical Research Consortium for the Participating Institutions, funding for the SDSS and SDSS-II has been provided by the Alfred P. Sloan Foundation, the Participating Institutions, the National Science Foundation, the US Department of Energy, the National Aeronautics and Space Administration, the MEXT, the Max Planck Society, and the Higher Education Funding Council for England. Pan-STARRS is supported by the National Aeronautics and Space Administration (NASA) under Grant No. NNX12AR65G and Grant No. NNX14AM74G, and the National Science Foundation (NSF) under Grant No. AST-1238877. The Gaia mission is a project of the European Space Agency (ESA) and funded by ESA member states and the European Union. Finally I would like to express my sincere gratitude to Professor Michael Richmond from the Rochester Institute of Technology for his invaluable guidance and suggestions during the final stages of this work. His expertise and insightful feedback greatly contributed to the overall clarity of this paper. We are truly grateful for his time, patience, and dedication in reviewing this work and providing valuable recommendations. His mentorship and support have been instrumental in shaping the final outcome of this paper.

\end{acknowledgments}
\clearpage

\bibliography{sample631}{}
\bibliographystyle{aasjournal}
\appendix
\section{The biases from the host galaxy subtraction}
\begin{figure}[!ht]
\centering
\gridline{\fig{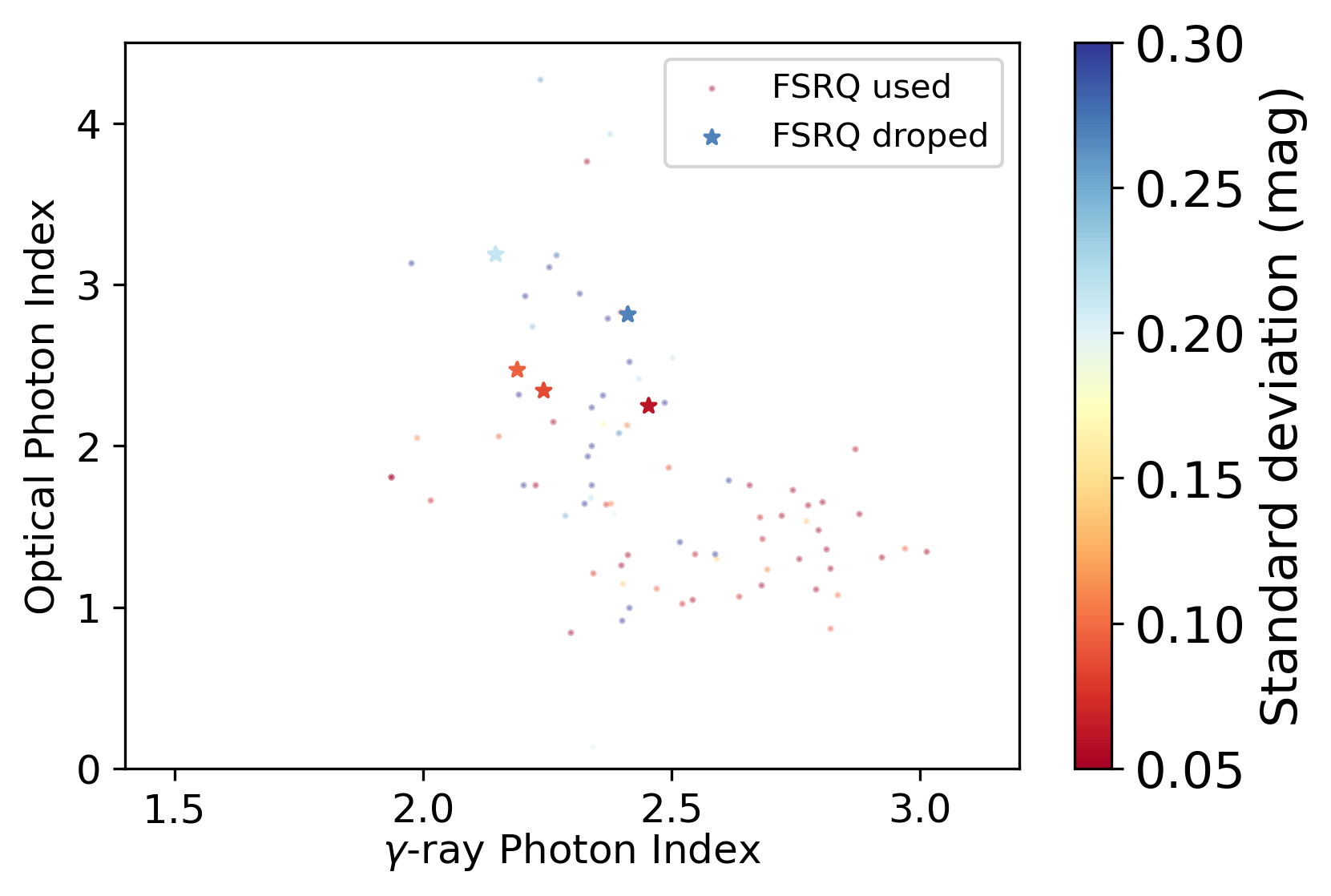}{0.5\textwidth}{FSRQs}
         \fig{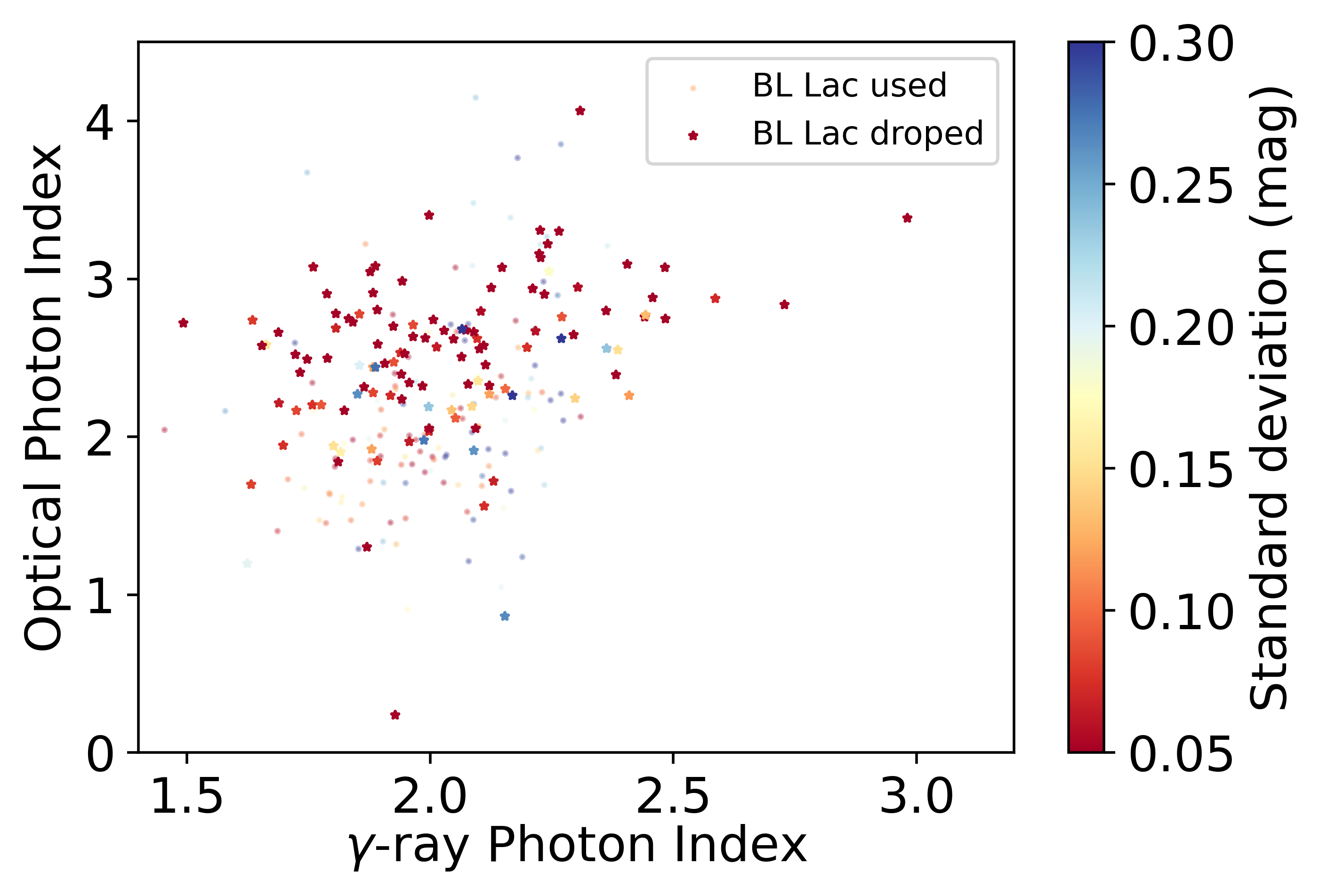}{0.5\textwidth}{Bl Lacs}
         }
\caption{Correlation between the $\gamma$-ray and optical photon indices for FSRQs and BL Lacs. The semi-transparent circular markers denote the FSRQs and BL Lacs included in this analysis, whereas the opaque star-shaped markers indicate those FSRQs and BL Lacs excluded due to potential contamination from host galaxy luminosity, which could lead to underestimation of variability (see section \ref{sec:Data selection}). The color coding of each symbol corresponds to the standard deviation of the optical photometry (light curve) measurements. }
\label{fig:all}
\end{figure}

Figure \ref{fig:all} shows the selection bias that may be introduced after the Data selection of section \ref{sec:Data selection}

It is evident from the figure that the photon index distribution of the excluded BL Lacs and FSRQs generally falls within the range of those utilized in this study. However, the mean optical photon index of these excluded BL Lacs and FSRQs is higher than that of the BL Lacs and FSRQs included at the same $\gamma$-ray photon indices. This provides substantial evidence that the optical bands of these blazars are likely contaminated by their host galaxies, resulting in a redder appearance compared to uncontaminated blazars. Several of the excluded BL Lacs exhibit $\gamma$-ray photon indices higher than those of the BL Lacs included in this study. Given that the excluded BL Lacs tend to have lower redshifts, and considering the pronounced evolution of blazars (see Fig. \ref{fig:gl}), it is entirely plausible that low-redshift BL Lacs have weaker jets and redder $\gamma$-ray spectra. Excluding these BL Lacs could potentially introduce a selection bias; however, as discussed in Section \ref{sec:Data selection}, distinguishing between the blazar component and the host galaxy contribution in optical band, as well as attributing their variability to either the blazar or the host galaxy, is not feasible. Given that these BL Lacs constitute less than three percent of the total BL Lac population, we believe their exclusion does not introduce a significant selection bias to the final conclusions.
\clearpage
\section{The biases from the standard deviation subtraction}
\begin{figure}[!ht]
\centering
\gridline{\fig{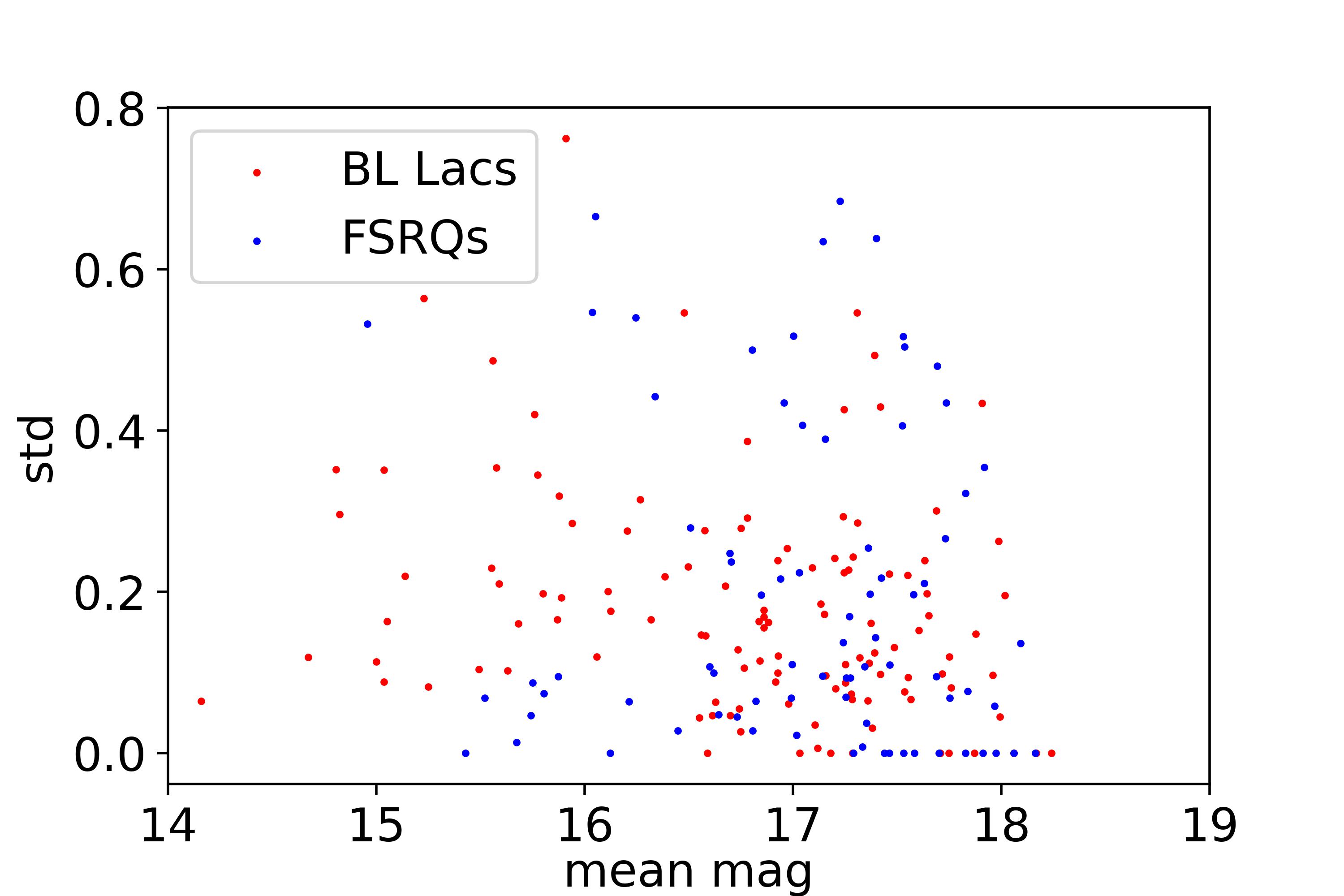}{0.5\textwidth}{a}
         \fig{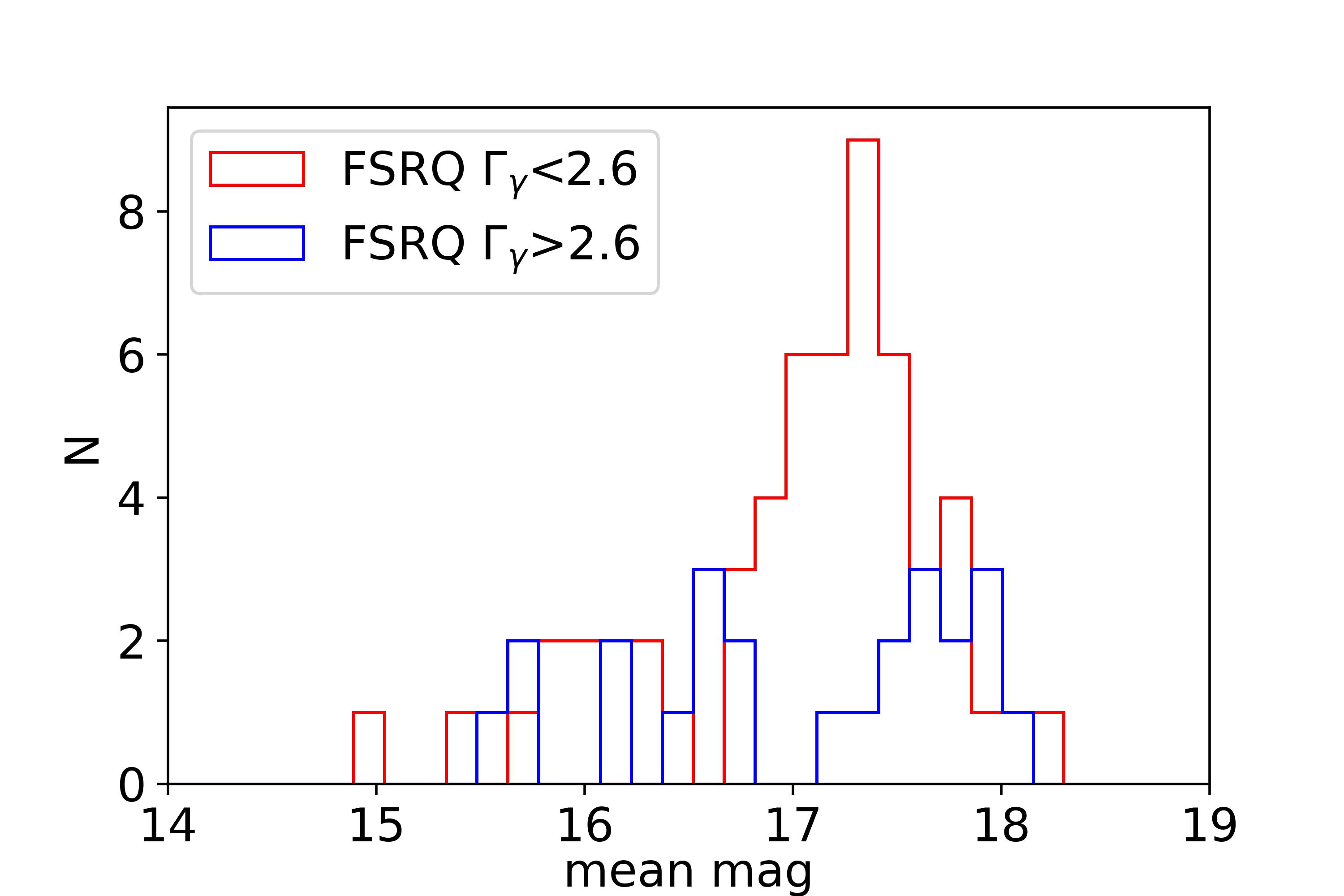}{0.5\textwidth}{b}
         }
\caption{a: Correlation between the mean mag and standard deviation for FSRQs and BL Lacs. b: Histogram of the mean mag of FSRQs with different $\gamma$-ray photon indices.}
\label{fig:a2}
\end{figure}
The limitations imposed by the systematic error in standard deviation subtraction are acknowledged, particularly in the context of darker objects where this method may lead to an underestimation of standard deviation. Despite this potential bias, our analysis indicates that this selection error remains consistent across various FSRQs with different $\gamma$-ray photon indices. This uniformity is evidenced by the similarity in the rank distribution of FSRQs with high $\gamma$-ray photon indices compared to those with standard $\gamma$-ray photon indices, as demonstrated by a KS-test yielding a p-value of 0.21. Consequently, we infer that the observed low variability in FSRQs with high $\gamma$-ray photon indices is not a result of the standard deviation subtraction process.

\begin{figure}[!ht]
\centering
\includegraphics[scale=1]{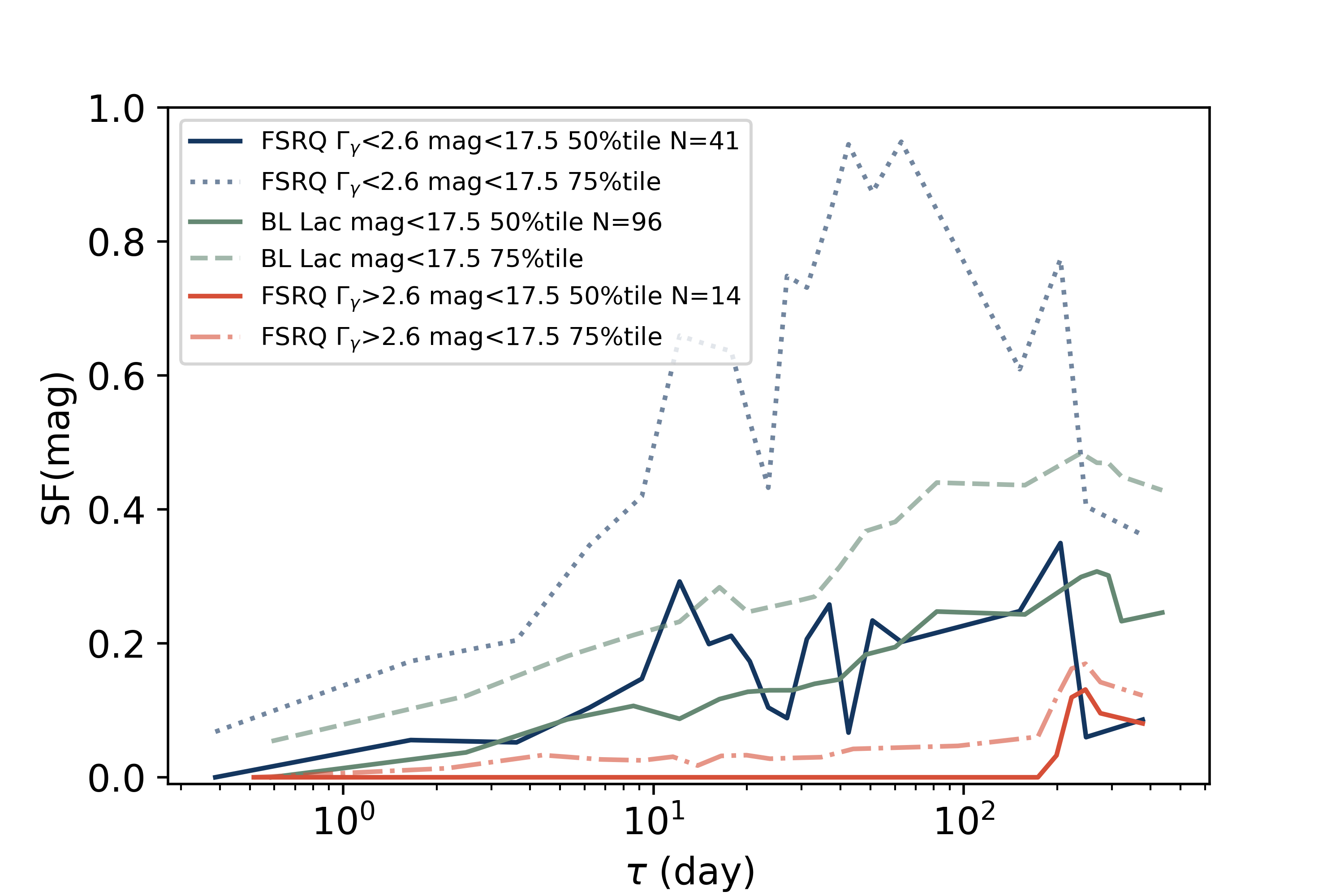}
\caption{Structure function of low $\gamma$-ray photon indices FSRQs with average Tomo-e Gozen magnitude brighter than 17.5 (blue; 7504 data pairs for each point), BL Lacs with average Tomo-e Gozen magnitude brighter than 17.5 (green; 32442 data pairs for each point) and high $\gamma$-ray photon indices FSRQs with average Tomo-e Gozen magnitude brighter than 17.5 (red; 2953 data pairs for each point).}
\label{fig:sfplmag}
\end{figure}

Also in our analysis, we have specifically plotted the structure function for blazars exhibiting an average Tomo-e Gozen magnitude brighter than 17.5, as depicted in Figure \ref{fig:sfplmag}. The observed behavior of the structure function in this figure closely mirrors that presented in Figure \ref{fig:sfpl}. Notably, the similarity in the patterns between these two figures substantiates the assertion that the biases potentially introduced by the standard deviation subtraction do not significantly impact our results.
\clearpage

\end{document}